\documentclass[12pt,preprint]{aastex}

\newcommand{\etal  }{{et al.} }
\newcommand{\msun}{\thinspace M_\odot}  
\newcommand{\rsun}{\thinspace R_\odot}  
\newcommand{\vect}[1]{\mbox{\boldmath$#1$}}

\newcommand{\rhoc}{\rho_{\rm c}}
\newcommand{\nc}{n_{\rm c}}

\newcommand{\ob}{\varepsilon_{\rm ob}}

\newcommand{\cm  }{\,{\rm cm}^{-3} } 
\newcommand{\jcm}{{\rm cm^2\,s^{-1}}}
\newcommand{\gjcm}{{\rm g\,cm^2\,s^{-1}}}
\newcommand{\dfrac}[2]{{\displaystyle \frac{#1}{#2}}  }
\newcommand{\km  }{\,{\rm km\, s^{-1}} } 
\newcommand{\kg  }{\,{\rm kG} } 
\newcommand{\betap }{\beta_{\rm p}}

\shorttitle{Driving Mechanism of Jets and Outflows}
\shortauthors{Machida  \etal 2006}

\begin{document}
\title{Driving Mechanism of Jets and Outflows in Star Formation Process}

\shortauthors{Machida  \etal 2006}

\author{Masahiro N. Machida\altaffilmark{1} and Shu-ichiro Inutsuka\altaffilmark{1}, and Tomoaki Matsumoto\altaffilmark{2}} 
\altaffiltext{1}{Department of Physics, Graduate School of Science, Kyoto University, Sakyo-ku, Kyoto 606-8502, Japan; machidam@scphys.kyoto-u.ac.jp, inutsuka@tap.scphys.kyoto-u.ac.jp}
\altaffiltext{2}{Faculty of Humanity and Environment, Hosei University, Fujimi, Chiyoda-ku, Tokyo 102-8160, Japan; matsu@i.hosei.ac.jp}

\begin{abstract}
The driving mechanism of jets and outflows in star formation process is studied using three-dimensional resistive MHD nested grid simulations. 
Starting with a Bonnor-Ebert isothermal cloud rotating in a uniform magnetic field, we calculated cloud evolution from the molecular cloud core ($\nc=10^4 \cm$, $r=4.6\times 10^4$\,AU) to the stellar core ($\nc=10^{22} \cm$, $r \sim 1 R_\odot$), where $\nc$ and $r$ denote the central density and radius of each object, respectively. 
In the collapsing cloud core, we found two distinct flows:
Low-velocity flows ($\sim$5$\km$) with a wide opening angle, driven from the adiabatic core when the central density exceeds $\nc \gtrsim 10^{12}\cm$, and high-velocity flows ($\sim$30$\km$) with good collimation, driven from the protostar when the central density exceeds $\nc \gtrsim 10^{21}\cm$.
High-velocity flows are enclosed by low-velocity flows after protostar formation.
The difference in the degree of collimation between the two flows is caused by the strength of the magnetic field and configuration of the magnetic field lines.
The magnetic field around an adiabatic core is strong and has an hourglass configuration; therefore, flows from the adiabatic core are driven mainly by the magnetocentrifugal mechanism and guided by the hourglass-like field lines.
In contrast, the magnetic field around the protostar is weak and has a straight configuration owing to Ohmic dissipation in the high-density gas region.
Therefore, flows from the protostar are driven mainly by the magnetic pressure gradient force and guided by straight field lines. 
Differing depth of the gravitational potential between the adiabatic core and the protostar cause the difference of the flow speed.
Low-velocity flows correspond to the observed molecular outflows, while high-velocity flows correspond to the observed optical jets.
We suggest that the outflow and the jet are driven by different cores, rather than that the outflow being entrained by the jet. 
\end{abstract}


\section{Introduction}
Since Snell discovered an outflow from a protostar in 1980 \citep{snell80}, over 300 outflows have been found in many star-forming regions \citep{wu04}.
Recently, outflows from brown dwarf \citep{whelan05} and O-type stars \citep{beuther05} were discovered.
These observations indicate that outflow is ubiquitous in the star formation process.
Observations also show that outflows have two or more distinct velocity components \citep[e.g.,][]{hirth97,pyo03,pyo05}.
Typically, an outflow is composed of a low-velocity component (LVC) with 10-50$\km$ and a high-velocity component (HVC) with $\sim$$100\km$ \citep{pyo05}.
Outflows have a collimation factor (i.e., length/width or major/minor radius of the outflow) of  $H/R = 3$ to $>20$  \citep{arce06}, where $H$ and $R$ are the height and radius of the outflow, respectively.  
There is a clear trend toward higher collimation at higher outflow velocity \citep{arce06}. 
Since outflows have various morphological and kinematical properties, they cannot be explained by a single-class  model.
The flows that originated from the protostar are typically classified into two types: molecular outflow (hereafter, outflow) observed mainly with CO molecules, and optical jet (hereafter jet) observed by optical emission \citep{arce06,pudritz06}.
Outflows observed by CO line emission exhibit a wide opening angle \citep[e.g.,][]{belloche02} and slow velocity  \citep[$10-50\km$;][]{richer00, arce06}. 
Jets observed by optical emission, however, exhibit good collimation \citep[e.g.,][]{hirano06} and high velocity  \citep[$100-500\km$;][]{arce06}.
Observations indicate that around each protostar, high-speed jets with a narrow opening angle are enclosed by a low-velocity outflow with a wide opening angle \citep[e.g.,][]{mundt83}. 

What mechanism drives the outflow and the jet in the star formation process? 
During star formation, neither radiation nor thermal pressure from the protostar can supply sufficient kinetic energy and momentum  for driving the jet and the outflow \citep{wu04,pudritz06}.
These flows (outflows and jets) are considered to be driven by release of the gravitational energy mediated by the Lorentz and centrifugal forces \citep[cf.][]{blandford82,shu94}.
However, we have not understood what drives these flows during star formation.
It is difficult to directly observe the driving points of the outflow and the jet, because they are embedded in the dense cloud at several AU from the protostar.
Furthermore,  we cannot analytically investigate the driving mechanisms of the jet and the outflow in detail, because they have a large spatial scale and density contrast, and these flows are influenced by closely related Lorentz, centrifugal, thermal pressure gradient, and gravitational forces.
Therefore, numerical simulation is the most effective tool for investigating the driving mechanisms of the jet and the outflow.

\citet{blandford82} showed that centrifugally driven outflow from the disk is possible if the poloidal component of the magnetic field makes an angle of less than 60$\degr$ (i.e., the disk wind model).
\citet{pudritz83,pudritz86} proposed that this models as the mechanism for the protostellar jet. 
Subsequently many numerical simulations \citep[e.g.,][]{kudoh97a,kudoh97b}, which showed that disk could drive the jet confirmed the disk-wind mechanism around the protostar \citep[see,][]{konigl00,pudritz06}.
\citet{shu94} proposed the X-wind model, in which the jet is driven in close proximity to  the protostar.
Both the disk wind and the X-wind are driven magnetocenrifugally from open magnetic field lines anchored on rapidly rotating circumstellar disks.
The difference between these models lies in where the field lines are anchored: near the radius of magnetospherical truncation on the disk for X-wind and over a wider range of disk radii for disk wind \citep{shang06}.
The jet speeds derived from  both X-wind and disk wind theories correspond to the Kepler speed at their driving point on the circumstellar disk.
Thus, these models cannot explain the two velocity components of jet and outflow.
Many people believe that outflow (i.e., slow speed and wide opening angle flow) is entrained by the jet (i.e., high speed and narrow opening angle flow), and that the jet is driven by either the disk wind or the X-wind \citep[e.g.,][]{arce06}.
However, almost all simulations, except for a few studies, calculate the evolution of the circumstellar disk (i.e., jet driving) after the protostar grows up sufficiently, and provide the idealized boundary, density profile of the disk, distribution of the magnetic field and velocity, accretion rate, etc.
After the protostar is formed, we cannot judge whether these configurations are realized. 
Therefore to acquire adequate information about the protostar and to understand the driving mechanism of the jet and the outflows, we should calculate  cloud evolution from the molecular cloud that forms the protostar.

Simulations of cloud evolution from the molecular cloud core to stellar core formation show that the low-velocity flow is driven from the adiabatic core formed before protostar formation, and that the high-velocity flow is driven from another core.
Assuming spherical symmetry, the evolution from molecular cloud  to stellar core has been investigated by many authors \citep[e.g.,][]{larson69,winkler80a,winkler80b,masunaga98,masunaga00}, and they determined the density, velocity, and temperature structures of the collapsing cloud and stellar core.
They have shown that the adiabatic core (or the first core) with a radius of $\sim$1\,AU forms when the central density reaches $\nc \simeq 10^{11}\cm$, because the gas becomes optically thick and the thermal pressure increases.
\citet{bate98} and \citet{whitehouse06} calculated the stellar core formation from the molecular cloud core using SPH simulations.
However, the outflow and the jet do not appear in their calculations because they ignored the magnetic effect.
The evolution of the magnetized cloud until the formation of the adiabatic core (hereafter, the first core) was investigated by \citet{machida04,machida05a,machida05b,machida06a}, \citet{hosking04}, \citet{matsu04}, \citet{ziegler05}, and \citet{fromang06}. 
The outflow appears after the first core formation in \citet{machida04,machida05b}, \citet{matsu04}, and \citet{fromang06}.
\citet{tomisaka98,tomisaka00,tomisaka02}, and \citet{banerjee06} investigated the evolution of the magnetized clouds up to protostar formation.
They calculated the cloud evolution from the molecular cloud ($\nc \simeq 10^2-10^6\cm$) to protostar formation ($\nc\simeq 10^{22}\cm$).
In their calculations, two distinct flows appeared: low-velocity flow ($v \simeq2$km\,s$^{-1}$) driven from the first core, and high-velocity flow ($ v \simeq 30$ km\,s$^{-1}$)  driven from the second core (i.e., the protostar).
 They expected that low-velocity flow would be observed as molecular outflow and high-velocity flow as an optical jet. 
They adopted ideal MHD approximation, which is valid in the low-density gas region ($n \lesssim 10^{12}\cm$); however, it is not valid in the high-density gas region ($n \gtrsim 10^{12}\cm$).
  \citet{nakano02} found significant magnetic flux loss for $10^{12} \cm \lesssim n \lesssim 10^{15}\cm$ by Ohmic dissipation.
Therefore,  \citet{tomisaka98,tomisaka00,tomisaka02} and \citet{banerjee06} overestimated the magnetic flux of the cloud, particularly in the high-density gas region.

 There are two predictions concerning driving outflows and jets: one is that the outflow is entrained by the jet, and another is that the outflow and jet are driven from different objects.
These predictions are mutually conflicting. 
In this paper, we calculate cloud evolution from the molecular cloud core ($n_c = 10^4\cm$, $r_c = 4.6 \times 10^4$\,AU) to stellar core formation ($n_c \simeq 10^{22} \cm$, $r_c \simeq 1 \rsun$) using three-dimensional resistive MHD nested grid method,  study the formation process of jets and outflows, and show the driving mechanisms of these flows.
The numerical method of our computations and our model framework are given in \S 2, and the numerical results are presented in \S 3.  
  We discuss the driving mechanism of the outflow and the jet in \S 4, and summarize our conclusions in \S 5.

\section{Model}
 Our initial settings are almost the same as those of \citet{machida06a,machida06b,machida07}.
To study cloud evolution, we use the three-dimensional resistive MHD nested grid code. 
We solve the resistive MHD equations including self-gravity:  
\begin{eqnarray} 
& \dfrac{\partial \rho}{\partial t}  + \nabla \cdot (\rho \vect{v}) = 0, & \\
& \rho \dfrac{\partial \vect{v}}{\partial t} 
    + \rho(\vect{v} \cdot \nabla)\vect{v} =
    - \nabla P - \dfrac{1}{4 \pi} \vect{B} \times (\nabla \times \vect{B})
    - \rho \nabla \phi, & 
\label{eq:eom} \\ 
& \dfrac{\partial \vect{B}}{\partial t} = 
   \nabla \times (\vect{v} \times \vect{B}) + \eta \nabla^2 \vect{B}, & 
\label{eq:reg}\\
& \nabla^2 \phi = 4 \pi G \rho, &
\end{eqnarray}
 where $\rho$, $\vect{v}$, $P$, $\vect{B} $, $\eta$, and $\phi$ denote the density, 
velocity, pressure, magnetic flux density, resistivity, and gravitational potential, respectively. 
The resistivity ($\eta$) in equation~(\ref{eq:reg}) is a function of density and  temperature \citep{nakano02}. 
For resistivity ($\eta$), we use the value adopted in \citet{machida06b,machida07}.
 To mimic the temperature evolution determined by \citet{masunaga00}, we adopt the piece-wise polytropic equation of state as
\begin{equation} 
P = \left\{
\begin{array}{ll}
 c_s^2 \rho & \rho < \rho_c, \\
 c_s^2 \rho_c \left( \frac{\rho}{\rho_c}\right)^{7/5} &\rho_c < \rho < \rho_d, \\
 c_s^2 \rho_c \left( \frac{\rho_d}{\rho_c}\right)^{7/5} \left( \frac{\rho}{\rho_d} \right)^{1.1}
 & \rho_d < \rho < \rho_e, \\
 c_s^2 \rho_c \left( \frac{\rho_d}{\rho_c}\right)^{7/5} \left( \frac{\rho_e}{\rho_d} \right)^{1.1}
 \left( \frac{\rho}{\rho_e}   \right)^{5/3}
 & \rho > \rho_e, 
\label{eq:eos}
\end{array}
\right.  
\end{equation}
 where $c_s = 190$\,m\,s$^{-1}$, 
$ \rho_c = 3.84 \times 10^{-13} \, \rm{g} \, \cm$ ($n_c = 10^{11} \cm$), 
$ \rho_d = 3.84 \times 10^{-8} \, \rm{g} \, \cm$  ($n_d =  10^{16} \cm$), and
$ \rho_e = 3.84 \times 10^{-3} \, \rm{g} \, \cm$  ($n_e = 10^{21} \cm$).
For convenience, we define ``the protostar formation epoch" as that at which the central density ($n_c$) reaches  $\nc = 10^{21} \cm$.
We also call the period for which $n_c < 10^{11}\cm$ ``the isothermal phase,"  the period for which $10^{11}\cm < \nc < 10^{16}\cm$ ``the adiabatic phase," the period for which $10^{16}\cm < \nc < 10^{21}\cm$ ``the second collapse phase," and the period for which $\nc > 10^{21}\cm$ ``the protostellar phase."

 In this paper, we adopt a spherical cloud with critical Bonnor-Ebert \citep{ebert55, bonnor56} density profile having $\rho_{c,0} = 3.841 \times 10^{-20} \, \rm{g} \, \cm$ ($n_{c,0} = 10^4\cm$) of the central (number) density as the initial condition.
 The critical radius for a Bonnor--Ebert sphere $R_c = 6.45\, c_s [4\pi G \rho_{BE}(0)]^{-1/2}$ corresponds to $ R_c = 4.58 \times 10^4$\,AU for our settings. 
 Initially, the cloud rotates rigidly ($\Omega_0$) around the $z$-axis and has a uniform magnetic field ($B_0 =17 $$\mu$G) parallel to the $z$-axis (or rotation axis).
 To promote contraction, we increase the density by 70\% from the critical Bonnor-Ebert sphere.

 The initial model is characterized by a single one non-dimensional parameter $\omega$.
This parameter is related to the cloud's rotation rate, and is defined using a central density $\rho_0$ as
\begin{equation}
\omega = \Omega_0/(4 \pi\,  G \, \rho_0  )^{1/2}.
\label{eq:omega}
\end{equation}
 The model parameters $\omega$,  magnetic field $B_0$,  angular velocity $\Omega_0$, total mass inside the critical radius $M_0$, and the ratio of the thermal $\alpha_0$,  rotational $\beta_0$,  and magnetic $\gamma_0$ energies to the gravitational energy,\footnote{
Representing the thermal, rotational, magnetic, and gravitational energies as $U$, $K$, $M$, and $W$, the relative factors against the gravitational energy are defined as $\alpha_0 = U/|W|$, $\beta_0 = K/|W|$, and $\gamma_0 = M/|W|$.
} are summarized in Table~\ref{table:init}.

 We adopt the nested grid method  \citep[for details, see][]{machida05a,machida06a} to obtain high spatial resolution near the center.
  Each level of a rectangular grid has the same number of cells ($ 64 \times 64 \times 32 $),  although the cell width $h(l)$ depends on the grid level $l$.
 The cell width is reduced by a factor of 1/2 as the grid level increases by 1 ($l \rightarrow l+1$).
 We assume mirror symmetry with respect to $z$ = 0.
 The highest level of a grid changes dynamically.
 We begin our calculations with four grid levels ($l=1$, 2, 3, 4).
 The box size of the initial finest grid $l=4$ is chosen to be $2 R_{\rm c}$, where $R_c$  denotes the radius of the critical Bonnor--Ebert sphere. 
 The coarsest grid $l=1$, therefore, has a box size of $2^4\, R_{\rm c}$. 
 A boundary condition is imposed at $r=2^4\, R_{\rm c}$, such that the magnetic field and ambient gas rotate at an angular velocity of $\Omega_0$ (for details, see \citealt{matsu04}).
  A new finer grid is generated whenever the minimum local  
$ \lambda _{\rm J} $ becomes smaller than $ 8\, h (l_{\rm max}) $. 
The maximum level of grids is restricted to $l_{\rm max} = 30$.
 Since the density is highest in the finest grid, the generation of a new grid ensures the Jeans condition of \citet{truelove97} with a margin of safety factor of two.
 We adopted the hyperbolic divergence $\vect{B}$ cleaning method of \citet{dedner02}.

\section{Results}
Assuming spherical symmetry, many authors investigated the evolution from molecular cloud to stellar core using radiation hydrodynamic calculations  \citep[e.g.,][]{larson69,winkler80a,winkler80b,masunaga98,masunaga00}. 
We briefly summarize the evolution of the collapsing cloud core, according to their calculations.
The molecular gas obeys the isothermal equation of state with a temperature of $\sim$10\,K until $\nc \simeq 5\times 10^{10}\cm$ (isothermal phase), then the cloud collapses almost adiabatically for $5\times 10^{10} \cm \lesssim \nc \lesssim 10^{16}\cm$ (adiabatic phase).
A quasi-static core (i.e., the first core) with $\sim$$10^{-2}\msun$ forms in the adiabatic phase.
Subsequently dissociation of molecular hydrogen when the density exceeds $\nc \gtrsim 10^{16}\cm$ triggers further gravitational collapse (i.e., second collapse).
Finally, the second core (i.e., protostar) with $\sim$$10^{-3} \msun$ forms at  $\nc \simeq 10^{21}\cm$.
The protostar increases its mass for the subsequent gas accretion phase.

In this study, we calculated the cloud evolution from the molecular cloud core with $\nc=10^{4}\cm$ to form a protostar with  $\nc \simeq 10^{23}\cm$.
\citet{machida04,machida05a,machida05b,machida06a} investigated cloud evolution in the isothermal and early adiabatic phases.
The evolution of the magnetic field and angular velocity in the collapsing cloud for $10^4 \cm \lesssim \nc \lesssim 10^{23}\cm$ was shown by \citet{machida07}.
Therefore, this paper focuses mainly on the cloud evolution  after the first core formation ($\nc \gtrsim 10^{11}\cm$) because we are interested in the driving mechanism of the outflow and the jet.
We parameterized angular velocity, magnetic field strength, and the ratio of thermal to gravitational energy of the initial cloud for 80 models.
Cloud evolutions are typically classified into three types when the outflow or jet appears in the collapsing cloud.
We also found that cloud evolution depends on the ratio of the angular velocity to the magnetic field ($\Omega_0/B_0$),  but does not depend on the ratio of thermal to  gravitational energy, especially after the first core formation  ($\nc \gtrsim 10^{11}\cm$).
Observations indicate that molecular cloud cores have large magnetic energies and small rotational energies \citep{crutcher99,caselli02}.
In these types of clouds, cloud evolution is sensitive to the initial angular velocity, as shown in \citet{machida05a,machida06a,machida07}.
In the following, we show only three typical models in which the initial clouds have the same magnetic field strengths and ratios of thermal to gravitational energy, but different angular velocities (slow, moderate, and rapid rotation rate).

\subsection{Cloud Evolution with Slow Rotation}
\subsubsection{First Core and Protostar Formation}
\begin{figure}
\begin{center}
\includegraphics[width=160mm]{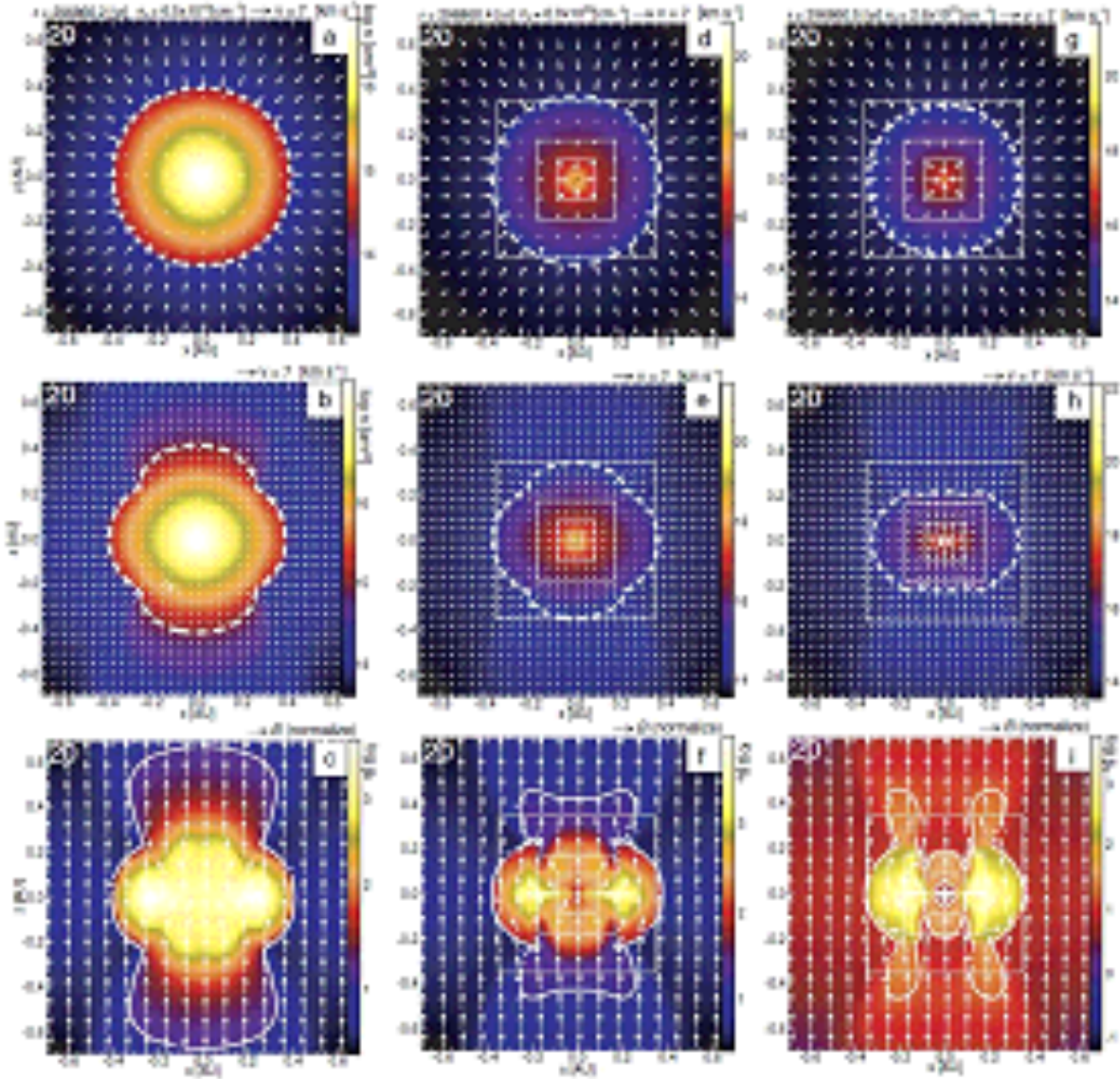}
\caption{
Time sequence of  model SR. 
Upper panels: Density (color-scale) and velocity distribution (arrows) on the cross-section in the $z=0$ plane ({\it a}, {\it d}, and {\it g}). 
Middle panels: Density (color-scale) and velocity distribution (arrows) on the cross-section in the $y=0$ plane ({\it b}, {\it e}, and {\it h}). 
Lower panels: Plasma beta (color-scale and contours) and magnetic field (arrows) on the cross-section in the $y=0$ plane ({\it c}, {\it f}, and {\it i}).
Panels from left to right are snapshots at the stages of 
 $n_c =6.5 \times 10^{16} \cm$ ($l=20$, panels {\it a}, {\it b}, and {\it c}), \ 
 $ 6.0 \times 10^{20} \cm$ ($l=20$; panels {\it d}, {\it e}, and {\it f}),\ and \ 
 $ 2.6 \times 10^{21} \cm$ ($l=20$; panels {\it g}, {\it h}, and {\it i}), respectively, 
where $l$ denotes the level of subgrid.
White-dotted lines in upper and middle panels show the shocked regions, which indicate the first core.
The elapsed time, density at the center of the cloud and arrow scale are denoted in each panel.
}
\label{fig:1}
\end{center}
\end{figure}

Figure~\ref{fig:1} shows the cloud evolution for model SR after the first core formation.
Model SR has a parameter of $\omega= 0.003$, which means that the initial cloud has a small rotation energy; the ratio of the rotational to gravitational energy is $\beta_0 = 3\times 10^{-5}$ (see Table~\ref{table:init}). 
Table~\ref{table:init} shows that, in model SR,  the rotational energy is much smaller than the thermal and magnetic energy in its initial state.

The density (false colors) and velocity (arrows) distribution around the first core are plotted in the upper and middle panels of Figure~\ref{fig:1}.
The plasma beta and magnetic fields around the first core are plotted in the lower panels of Figure~\ref{fig:1}.
We plotted shocked regions as dotted lines in the upper and middle panels in Figure~\ref{fig:1}. 
Since the first core is surrounded by the shock layer \citep{masunaga00},  dotted lines indicate the first core.
For model SR, the first core forms when the central density reaches $\nc \simeq 10^{14}\cm$.
The first core at its formation epoch has a mass of $5.1\times 10^{-3}\msun$ and a radius of $R\simeq 0.71$\,AU, which is equivalent to or slightly larger than that expected by one-dimensional radiation hydrodynamic calculations \citep{masunaga00}.
\citet{matsu03} showed that a larger first core forms in a rapidly rotating cloud  because the centrifugal force suppresses the cloud collapse and the shock occurs in the earlier adiabatic phase.
However, the effect of centrifugal force at the first core formation epoch is small in model SR because a slowly rotating cloud is adopted as the initial state.
The magnetic effect slightly increases the mass and size of the first core, as shown in \citet{machida05a}.

\begin{figure}
\begin{center}
\includegraphics[width=160mm]{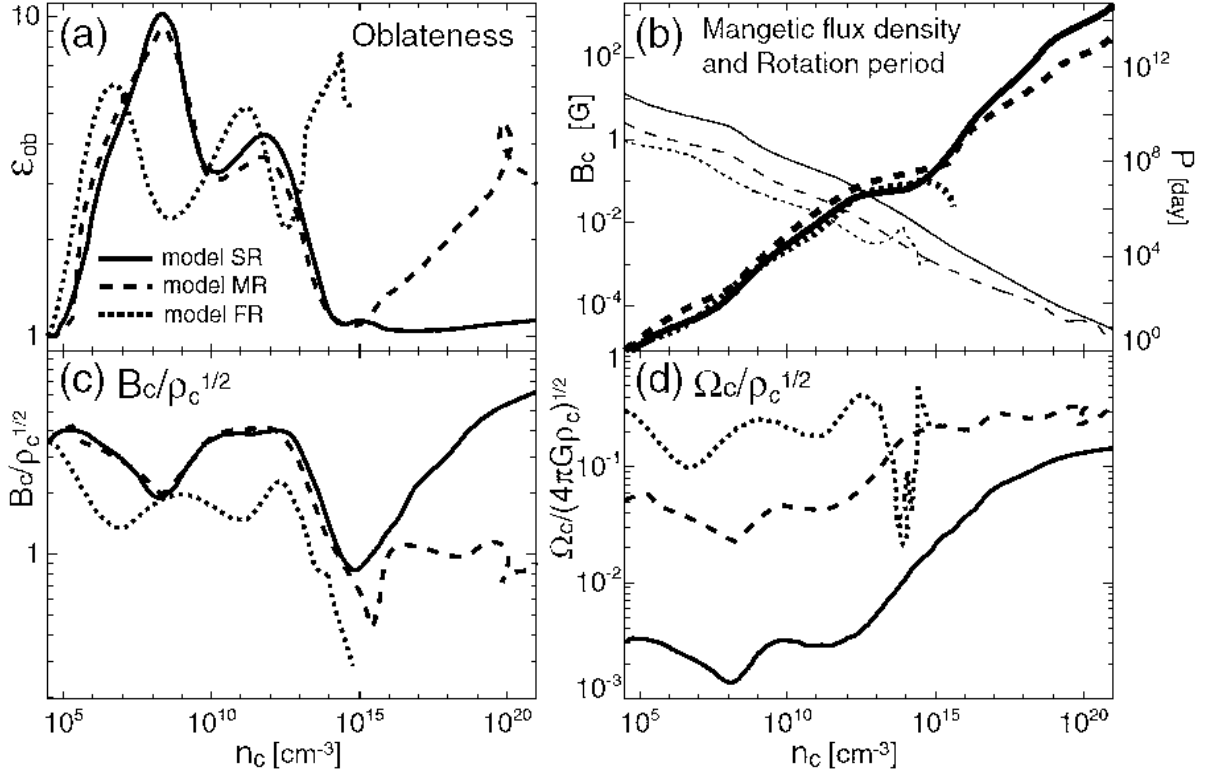}
\caption{
(a) Oblateness ($\ob$) within $\rho<0.1 \rho_{\rm c}$, (b) Thick lines: Magnetic flux density ($B_{\rm c}$; left axis). Thin lines: rotation period ($P$; right axis),  (c)  Magnetic field normalized by the square root of the density ($B_{\rm c}/\rhoc^{1/2}$), and (d): Angular velocity normalized by the free-fall timescale  [$\Omega_c/(4\pi G \rhoc)^{1/2}$] at the center of the cloud against the central number density ($\nc$) for models SR, MR, and FR.
}
\label{fig:2}
\end{center}
\end{figure}

Figures~\ref{fig:1}{\it a} and {\it b} show the density and velocity distribution on $z=0$ and $y=0$ cut planes at $\nc = 6.5 \times 10^{16}\cm$.
These panels indicate that the first core has a nearly spherical shape.
To evaluate the core shape, we define the oblateness as  $ \ob \, \equiv \, (h _l h _s) ^{1/2} / h _z $, where $ h _l $, $ h _s $, and $ h _z$ are the major axis, minor axis, and $ z $-axis, respectively, derived from the moment of inertia for the high-density gas of $ \rho \, \ge \, 0.1  \rho _{\rm c}  $ according to Matsumoto \& Hanawa (1999).
Figure~\ref{fig:2}{\it a} shows the evolution of the oblateness as a solid line.
The oblateness increases as the cloud collapses in the isothermal phase ($\nc \lesssim 10^{14} \cm$), and has a peak of $\ob \simeq 10$ (i.e., the ratio of the radial to vertical scale is about 10) at $\nc \simeq 2 \times 10^{8} \cm$.
Therefore, the central region has a disk-like structure at this epoch.
Since the cloud rotates slowly, the disk is formed mainly by the Lorentz force.
Note that the magnetic energy is comparable to both the thermal and gravitational energy  at the initial state (see Table~\ref{table:init}).
The oblateness oscillates around $\ob \simeq 10$ for $2\times 10^8 \cm \lesssim \nc \lesssim 10^{12}\cm$, and then begins to decrease (Fig.~\ref{fig:1}{\it a}).
This decrease is caused by an increase in thermal energy.
When the gas density reaches $\nc \simeq 10^{11}\cm$, the cloud collapses adiabatically and thermal pressure increases.
The oblateness becomes $\ob\simeq 1$ at the first core formation epoch ($\nc \simeq 10^{14}\cm$), which indicates that the first core has a nearly spherical shape, as shown in Figure~\ref{fig:1}{\it a} and {\it b}.

Figure~\ref{fig:1}{\it c} shows the plasma beta ($\betap\equiv B^2/8\pi c_s^2 \rho$) around the first core on the $y=0$ cut plane.
This panel shows that the magnetic energy inside the first core is extremely small ($\betap \simeq 10-10^4$), while that outside the first core is comparable to the thermal energy ($\betap < 1$).
As shown in \citet{machida06b,machida07}, the magnetic field is largely removed from the cloud core by Ohmic dissipation for $10^{12}\cm \lesssim \nc \lesssim 10^{15}\cm$.
Figure~\ref{fig:2}{\it b} shows the evolution of the magnetic flux density at the center of the cloud ($B_{\rm c}$) for model SR (solid line).
The growth rate of the magnetic flux density becomes small for $10^{12}\cm \lesssim \nc \lesssim 10^{15}\cm$.
The solid line in Figure~\ref{fig:2}{\it c} shows the magnetic flux density normalized by the square root of the gas density at the center of the cloud ($B_{\rm c}/\rhoc^{1/2}$) for model SR.
In the ideal MHD regime, the evolution of  $B_{\rm c}/\rhoc^{1/2}$ depends on the geometry of the collapse.
This value, $B_{\rm c}/\rhoc^{1/2}$, remains constant after a thin disk forms because the magnetic field increases in proportional to the square root of the density when the disk-like cloud collapses  ($B \propto \rho^{1/2}$; \citealt{scott80}).
On the other hand, $B_{\rm c}/\rhoc^{1/2}$ increases in  proportion to $\rho^{1/6}$ when the cloud collapses spherically \citep[for details, see][]{machida06a}.
Thus, a rapid drop of  $B_{\rm c}/\rhoc^{1/2}$ for $10^{12}\cm \lesssim \nc \lesssim 10^{15}\cm$ indicates the removal of the magnetic field from the center of the cloud by Ohmic dissipation.
\citet{nakano02} showed that the magnetic field is largely removed by  Ohmic dissipation for $10^{12}\lesssim \nc \lesssim 10^{16}\cm$.
Since the first core forms at $\nc \simeq 10^{12}\cm$ in this model, the first core is composed of the gas from which the magnetic field is already removed. 
Therefore, the first core has a large plasma beta (or weak magnetic field strength), as shown in Figure~\ref{fig:1}{\it c}.

Figures~\ref{fig:1}{\it d} and {\it e} show the density and velocity distributions around the first core at $\nc = 6 \times 10^{20}\cm$ (second collapse phase).
The cloud can collapse again for $10^{16}\cm \lesssim \nc \lesssim 10^{21}\cm$ owing to dissociation of molecular hydrogen (i.e., second collapse; \citealt{masunaga00}).
The arrows in Figure~\ref{fig:1}{\it d} and {\it e} indicate that the cloud collapses intensely around its center.
The size of first core at this epoch is slightly smaller than that at the epoch of Figure~\ref{fig:1}{\it a} and {\it b}.
However, the first core has a nearly spherical shape ($\ob\simeq 1$ at $\nc=6\times 10^{20}$ in Fig.~\ref{fig:2}{\it a}) because the anisotropic Lorentz and centrifugal forces are weaker  than the isotropic forces of the gravitational and thermal pressure gradient forces in this phase.
The Lorentz force becomes weak due to magnetic dissipation, and an initially small rotation rate produces a weaker centrifugal force.

Figure~\ref{fig:2}{\it d} shows the angular velocity normalized by the free-fall timescale at the center of the cloud, whose value corresponds to the ratio of the rotational to gravitational energy \citep{machida05a}.
When this value reaches $\Omega_{\rm c}/(4\pi G\rhoc)^{1/2} \simeq 0.2$, the rotational energy becomes comparable to the gravitational energy, and cloud rotation begins to affect cloud evolution \citep{matsu03,machida05a,machida06a}.
In model SR, the effect of cloud rotation is significantly small in the second collapse phase ($10^{16} \cm \lesssim \nc \lesssim 10^{20}\cm$) because $\Omega_{\rm c}/(4\pi G\rhoc)^{1/2} \ll 0.2$.
The magnetic field is again important for cloud evolution after the second collapse phase because Ohmic dissipation becomes less effective, and the magnetic field can be amplified. 
Figure~\ref{fig:2}{\it c} shows that $B_{c}/\rhoc^{1/2}$ has a minimum at $\nc \simeq 10^{15}\cm$, and then continues to increase for $\nc \gtrsim 10^{15}\cm$.
However, the plasma beta inside the first core reaches only $\betap \sim 100$ at $\nc \simeq 10^{21}\cm$ (Fig.\ref{fig:1}{\it f}), and therefore, the magnetic energy is much smaller than the thermal energy, even in the second collapse phase.

Figures~\ref{fig:1}{\it g} and {\it h} show the density and velocity distributions at $\nc = 2.6 \times 10^{21}\cm$ (the protostar formation epoch).
In our calculations, the first core (i.e., shock layer) does not disappear until the calculation ends.
The second core (i.e., protostar; \citealt{masunaga00}) forms inside the first core in Figures~\ref{fig:1}{\it g} and {\it h}.
Since both the magnetic field and angular velocity continue to increase in the second collapse phase, the first core is slightly flattened both by Lorentz and centrifugal forces at the protostar formation epoch, as shown in Figure~\ref{fig:1}{\it h}.
However, the protostar has an oblateness $\ob=1.2$ (Fig.~\ref{fig:2}{\it a}) at its formation epoch ($\nc = 10^{21}\cm$), and therefore, has a nearly spherical shape.
The magnetic field and rotation period reach $B = 2.18\kg$ and $P=3$\,days at the protostar formation epoch.
The plasma beta around the protostar  (or inside the first core) becomes $\betap \simeq 10-10^3$ (Fig.~\ref{fig:1}{\it i}).
After the protostar formation ($\nc \gtrsim 10^{21}\cm$), the plasma beta becomes large ($\betap \simeq 10^3-10^4$) because the thermal energy becomes large [see eq.~(\ref{eq:eos})].
The angular velocity normalized by the free-fall timescale, however, reaches $\Omega_{\rm c}/(4\pi G \rhoc)^{1/2} \simeq 0.2$ at the protostar formation epoch (Fig.~\ref{fig:2}{\it d}); then, the cloud begins to rotate rapidly, and the rotation significantly affects protostar evolution.

\subsubsection{Jet Formation}
\begin{figure}
\begin{center}
\vspace{-5mm}
\includegraphics[width=150mm]{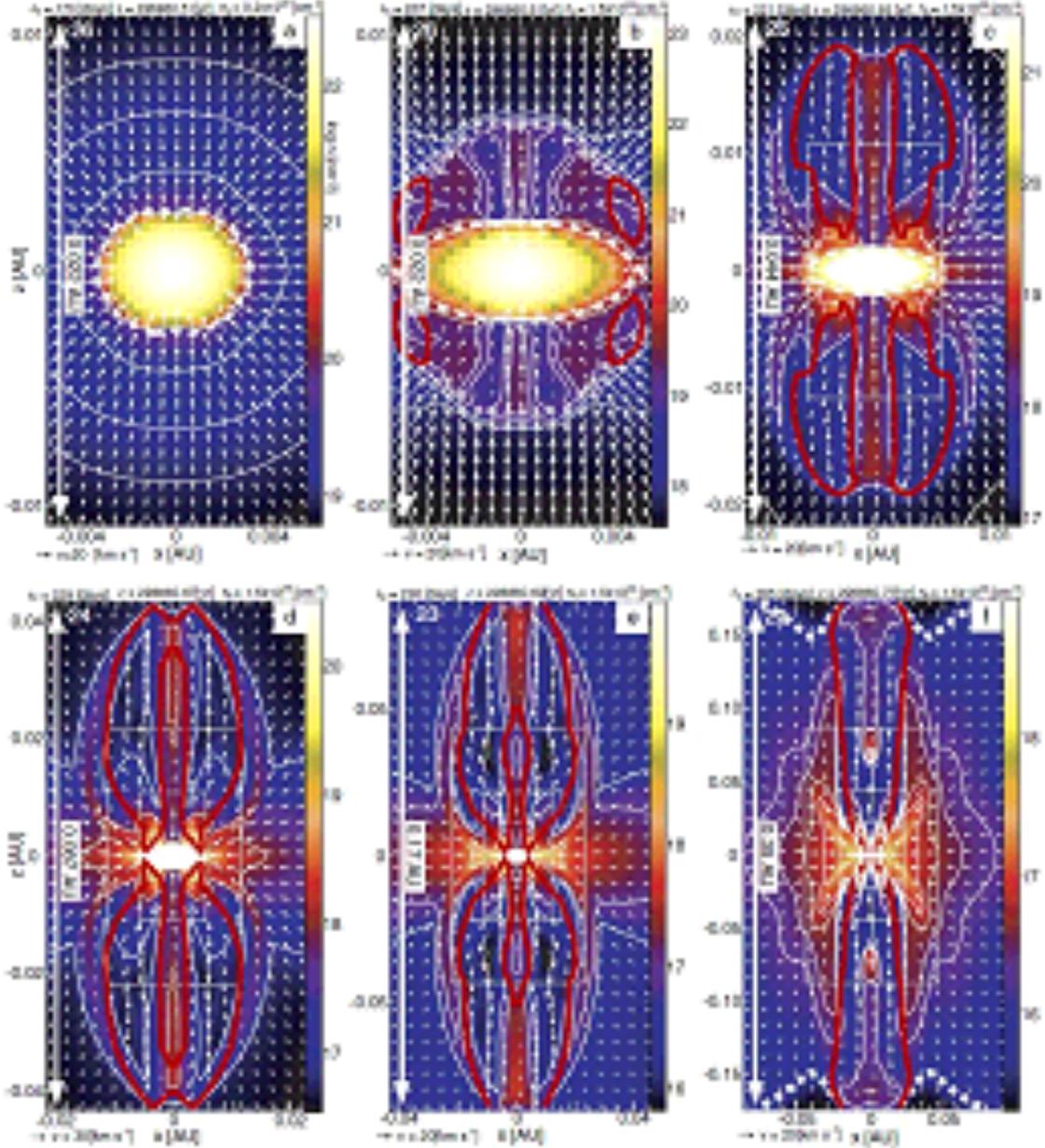}
\caption{
Time sequence of model SR after the protostar formation epoch .
The density (color-scale) and velocity distribution (arrows) on the cross-section in the $y=0$ plane are plotted in each panel.
Panels (a) through (f) are snapshots at the stages of 
(a) $ t_{\rm c} =170 $\,days ($ l=26 $), \ (b) 207\,days ($ l=26 $), \ (c) 221\,days ($ l=25 $), \ (d) 229\,days ($ l=24 $), \ (e) 236\,days ($ l=23 $), \ and \ (f) 265\,days ($l=22$), where $t_{\rm c}$ denotes the elapsed time after protostar formation ($\nc = 10^{21}\cm$).
Red thick lines mean the border between the jet and the accretion flow (i.e., contour of $v_z$ = 0).
White dotted lines in panels {\it a}-{\it c} indicate the second core (i.e., protostar), while white dotted line in panel {\it f} indicate the first core.
The elapsed time after protostar formation ($t_{\rm c}$), the elapsed time from the initial ($t$), gas density at the center of the cloud ($\nc$), arrow scale, and grid scale are denoted in each panel.
The level of the subgrid is shown in the upper left corner.
}
\label{fig:3}
\end{center}
\end{figure}

Figure~\ref{fig:3} shows cloud evolution after the protostar formation epoch ($\nc > 10^{21}\cm$).
Figures~\ref{fig:3}{\it a}--{\it c} show the protostar (i.e., second core) surrounded by the shock layer as a dotted line.
Figure~\ref{fig:3}{\it a} shows the density and velocity distribution around the protostar 170\,days after the protostar formation epoch.
The protostar has a mass of $5.1 \times 10^{-4}\msun$ and a radius of $R\simeq 0.75\rsun$ at its formation epoch ($\nc \simeq 10^{21}\cm$).
The arrows in Figure~\ref{fig:3}{\it a} show that the gas around the protostar accretes spherically onto the protostar.
Figure~\ref{fig:3}{\it b} shows the protostar 207\,days after its formation epoch.
Figures~\ref{fig:3}{\it a} and {\it b} show that the protostar becomes flat with time because the magnetic field and cloud rotation are amplified and affect protostar evolution, as shown in Figures~\ref{fig:2}{\it c} and {\it d}.
Two shocks (outer and inner) are seen in Figure~\ref{fig:3}{\it b}.
The outer shocks are located at $z\simeq \pm 6 \times 10^{-3}$\,AU, while the inner shocks corresponding to the protostar are located at $z\simeq \pm 2\times 10^{-3}$\,AU.
As shown in Figure~\ref{fig:3}{\it c},  the accretion speed inside the outer shock ($\vert v\vert \simeq 20\km$ in the region of $z<\vert2\times 10^{-3} \vert$\,AU) is slower than that outside the outer shock ($\vert v\vert \simeq 5\km$ in the region of $z>\vert2\times 10^{-3} \vert$\,AU).
This structure of the shock and distribution of the density and velocity, shown in Figure~\ref{fig:3}{\it b}, is similar to ``the magnetic bubble"  in \citet{tomisaka02}.
He showed that the magnetic pressure is amplified by the twisted magnetic field lines, and a bubble-like structure is formed in the weakly magnetized collapsing cloud.

In model SR, the jet appears 215\,days after the protostar formation epoch
\footnote{
For convenience, this paper refers to the flow driven from the first core as ``the outflow", and the flow driven from the second core (or protostar) as ``the jet."}.
Figure~\ref{fig:3}{\it c} shows that the gas flows out from the central region  along the vertical axis.
The red lines in Figures~\ref{fig:3}{\it b}-{\it f} indicate the border between the inflow and the outflow.
Inside the red lines, the gas flows out from the central region ($v_z \gtrless 0 $ for $z\gtrless0$), and outside the redlines, it flows into the central region ($v_z \lessgtr 0 $ for $z\gtrless0$).
The jet has a maximum speed of $v_{\rm jet} \simeq 17\km$ at the same epoch as in Figure~\ref{fig:3}{\it c}.
The jet has an elongated structure and  height-to-radius ratio of $H/R \approx 3$ in Figure~\ref{fig:3}{\it c}.
The height-to-radius ratio increases with time.
Figures~\ref{fig:3}{\it d} and {\it e} show the jet extending in the vertical direction keeping almost the same horizontal scale ($R=0.012$\,AU in Fig.~\ref{fig:3}{\it d} and $0.018$\,AU in Fig.~\ref{fig:3}{\it e}).
The jet extends up to $H=0.043$\,AU with a maximum speed of $v_{\rm jet} \simeq 22\km$ at $t_c =229$ days (Fig.~\ref{fig:3}{\it d}), while it extends up to $H=0.1$\,AU with $v_{\rm jet}\simeq 26\km$ at $t_{\rm c}=226$ days (Fig.~\ref{fig:3}{\it e}), where $t_{\rm c}$ is the elapsed time from the protostar formation epoch ($\nc = 10^{21}\cm$).
Figure~\ref{fig:3}{\it f} shows the structure of the jet at the end of the calculation.
Note the butterfly-like density distribution (white-density contour), which is caused by the strong mass ejection from the protostar.
The jet has a well-collimated structure at this epoch.
The height-to-radius ratio of the jet reaches $H/R\simeq 10$.
In Figure~\ref{fig:3}{\it f}, the jet penetrates the first core, shown as a thick broken line, and extends up to the outside of the first core.
At the end of the calculation, the jet reaches $0.2$\,AU with a maximum speed of $v_{\rm jet}\simeq 32\km$.

\begin{figure}
\begin{center}
\includegraphics[width=160mm]{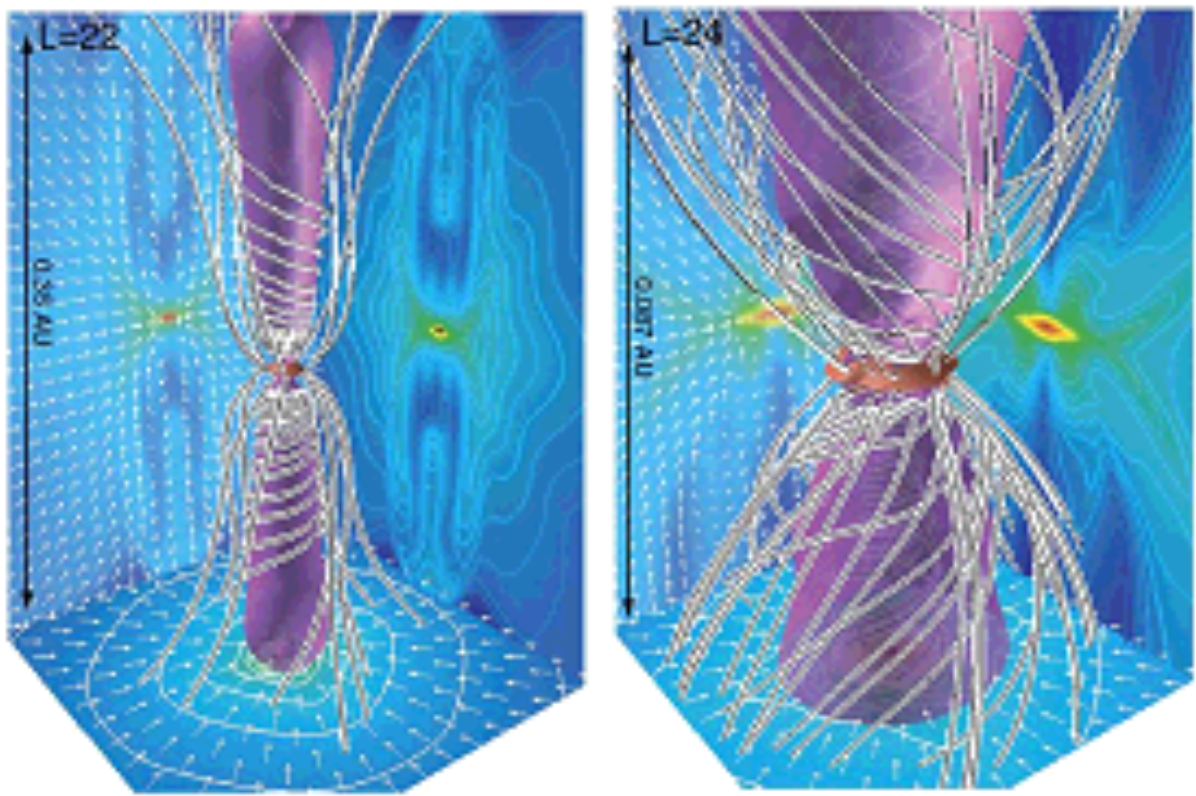}
\caption{
Bird's-eye view of model SR at the same epoch as Fig.~\ref{fig:3}{\it f}, but with different scales of $L=0.35$\,AU (left), and $L=0.087$\,AU (right), where $L$ means the grid scale.
The structure of high-density region ($\rho > 0.1\rho_{\rm c}$; red iso-density surface),  and magnetic field lines (black-and-white streamlines) are plotted in each panel.
The structure of the jet is shown by purple iso-velocity surface in which the gas is outflowing from the center.  
The density contours (false color and contour lines), velocity vectors (thin arrows) on the mid-plane of $x=0$, $y=0$, and $z=$0 are, respectively, projected in each wall surface.
}
\label{fig:4}
\end{center}
\end{figure}

Figure~\ref{fig:4} shows the shape of the jet (purple iso-velocity surface) and magnetic field lines (stream lines) at the same epoch as Figure~\ref{fig:3}{\it f} but with a different grid size (left: $l=22$, right: $l=24$).
Figure~\ref{fig:4} left panel confirms that the jet has a well-collimated structure showing that the jet is strongly coiled by the magnetic field lines, with the bow shocks clearly visible near the upper and lower boundaries.
Figure~\ref{fig:3} (right panel) provides a close-up view of the left panel, indicating that the jet is driven from the protostar, which is shown by a red iso-density surface.
We can see ripping density contours on the projected wall caused by strong mass ejection  \citep{machida06b}.

At the end of the calculation, the protostar in model SR has a mass of $4.04\times10^{-3}\msun$ and a radius of $1.07\rsun$, while the mass flowing out from the protostar is $M_{\rm jet}=5.23\times 10^{-5} \msun$.
At the same epoch, the protostar and jet have the angular momenta of $J_{\rm core} = 4.23\times 10^{47}\gjcm$ and $J_{\rm jet} = 1.55 \times 10^{46}\gjcm$, respectively.
Therefore, the specific angular momenta of the protostar and jet are $j_{\rm core} = 5.25\times 10^{16}\jcm$ and $j_{\rm jet} = 1.49 \times 10^{17}\cm$, respectively.
This indicates that the jet largely removes the angular momentum from the protostar.
\citet{machida07} showed that, at its formation epoch, the protostar has a short rotation period compared with the observation.
The jet may be one of the mechanisms for the angular momentum transfer after protostar formation.

\subsection{Cloud Evolution with  Moderate Rotation}
\subsubsection{First Core and Protostar Formation}
\begin{figure}
\begin{center}
\includegraphics[width=160mm]{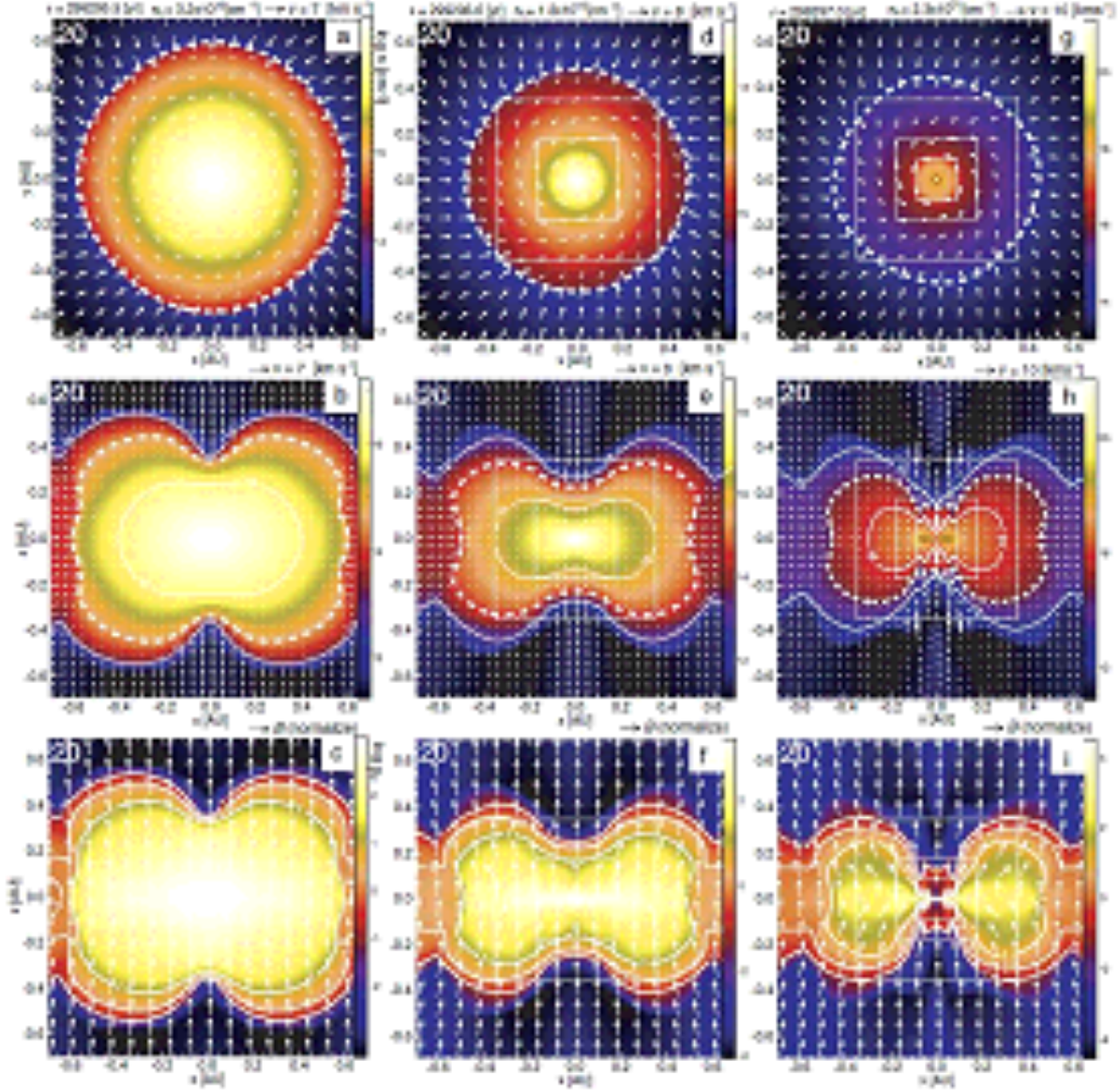}
\caption{
As same as Fig.~\ref{fig:2} but for model MR.
Panels from left to right  are snapshots at the stages of 
 $n_c =3.2 \times 10^{16} \cm$ ($l=20$, panels {\it a}, {\it b}, and {\it c}), \ 
 $ 1.2 \times 10^{18} \cm$ ($l=20$; panels {\it d}, {\it e}, and {\it f}),\ and \ 
 $ 2.3 \times 10^{21} \cm$ ($l=20$; panels {\it g}, {\it h}, and {\it i}), respectively. 
The white-dotted lines in upper and middle panels mean the first core, while the red-dotted lines in panels {\it g} and {\it h} indicate the protostar.
}
\label{fig:5}
\end{center}
\end{figure}

Figure~\ref{fig:5} shows the cloud evolution for model MR after first core formation.
Model MR has a parameter of $\omega= 0.03$.
The ratio of the rotational to gravitational energy for model MR is $\beta_0 = 3\times 10^{-3}$, which is 100 times larger than that for model SR at its initial state (see Table~\ref{table:init}).
In model MR, the first core forms at $\nc \simeq 10^{14}\cm$, and has a mass of $M\sim0.01\msun$ and a radius of $R\sim 1$\,AU.
Figures~\ref{fig:5}{\it a} and \ref{fig:1}{\it a} show that the first core of model MR is larger than that of model SR, because the centrifugal force for model MR is stronger.
The angular velocity, normalized by free-fall timescale at the center of the cloud, saturates $\Omega_c/(4\pi G \rhoc)^{1/2} \simeq 0.2$ for $\nc \gtrsim 10^{14}\cm$ (Fig.~\ref{fig:2}{\it d}).
This indicates that cloud rotation significantly affects cloud evolution after the central density reaches  $\nc \simeq 10^{14}\cm$.
Due to the rotation (or the centrifugal force), the first core for model SR, has a more oblate structure than that for model SR, as shown in Figure~\ref{fig:5}{\it b}.
The evolution of the oblateness for model MR is almost the same as that for model SR for $\nc \lesssim 10^{14}\cm$.
On the other hand, the oblateness increases after the central density reaches $\nc \simeq 10^{15}\cm$, while it maintains $\ob\simeq 1$ for $\nc \gtrsim 10^{14}\cm$ in model SR.
Figure~\ref{fig:5}{\it c} shows the plasma beta around the first core at $\nc=3.2\times10^{16}\cm$.
The plasma beta for model MR is as large as that for model SR ($\betap=10-10^{3}$) at the first core formation epoch (Fig.~\ref{fig:5}{\it c}). 
Thus, the magnetic field slightly affects cloud evolution inside the first core at this epoch.

Figures~\ref{fig:5}{\it d} and {\it e} show the density and velocity distribution at $\nc = 1.2 \times 10^{18}\cm$ (the second collapse phase).
The arrows in Figure~\ref{fig:5}{\it d} show that the cloud collapses while rotating: however, it collapses only in the radial direction for model SR (Fig.~\ref{fig:1}).
The azimuthal component of the velocity ($v_{\phi}$) inside the first core is comparable to or larger than the radial one ($v_{\rm r}$) in model MR, while the azimuthal component of the velocity is much smaller than the radial one in model SR.
Since the gas cloud inside the first core rotates rapidly, the central region increases its oblateness as the cloud collapses (Fig.~\ref{fig:2}{\it a}).
Therefore, the central region  has a more oblate or disk-like structure,  similar to the first core shown in \citet{saigo06}, in which they investigated the equilibrium state of the rotating first cores.
In general, since the angular momentum is transferred by magnetic braking in the magnetized clouds \citep{basu94}, the cloud rotation slows with time.
In this model, however, the cloud rotation hardly slows because the magnetic energy around the first core is too small to effectively work the magnetic braking.
The small energy of the magnetic field is realized by Ohmic dissipation.
Therefore, the cloud around the first core has a large plasma beta ($\betap>10$; Fig.~\ref{fig:5}{\it c}).
Figure~\ref{fig:2}{\it c} compares the magnetic fields normalized by the density at the center of the cloud  ($B_{\rm c}/\rhoc^{1/2}$) for model MR (broken line), which are smaller than those for model SR (solid line)  because for model MR, the cloud collapses, maintaining the disk-like structure, while that in model SR collapses spherically. 
The growth rate of the magnetic field depends on the geometry of the collapse and is much larger for a spherical collapse than for a disk-like collapse \citep{machida05a,machida06a}.
As a result, the protostar has a magnetic field of $B=0.4\kg$ for model MR which is five times smaller than that for model SR at its formation epoch ($\nc=10^{21}\cm$).

Figures~\ref{fig:5}{\it g} and {\it h} show the density and velocity distribution at $\nc = 2.3 \times 10^{21}\cm$ (the protostar formation epoch).
In each panel, the protostar (i.e., the second core) is represented by a red-dotted line.
In model MR, the protostar forms at $\nc \simeq 2\times 10^{20}\cm$, and has a mass of $M\simeq 3.6\times 10^{-4}\msun$ and a radius of $R=1.5\rsun$.
As shown in Figure~\ref{fig:5}{\it h}, at the protostar formation epoch, the first core sags downward in the center in a concave structure similar to model C in \citet{saigo06}.
This concave structure is considered to be caused by rapid rotation and an increase in the density of the first core \citep[for details, see][]{saigo06}.
As shown in Figure~\ref{fig:5}{\it h}, since the gas density falls below $\nc \lesssim 10^{12}\cm$ in the regions just above and below the first core at the center of the cloud ($\vert x \vert < 0.1$\,AU), Ohmic dissipation is less effective in these regions.
Note that  Ohmic dissipation is effective for $10^{12}\cm \lesssim \nc \lesssim 10^{15}\cm$ \citep{nakano02}.
Therefore, the magnetic field is strong around the protostar ($\betap\simeq 0.1$).
Since the central region becomes increasingly thinner and has an increasingly concave structure over time, the magnetic field around the protostar barely dissipates through Ohmic dissipation.
Therefore in model MR, the torsional alfv\'en wave generated by the rotation of the protostar can be transferred outside without dissipation. 
In model SR, however, the magnetic field dissipates and the plasma beta maintains a high value ($\betap\simeq10-10^3$) around the protostar, because the cloud collapses spherically with slow rotation, and Ohmic dissipation is effective around the protostar.
Figure~\ref{fig:5}{\it h} shows that the outflows are driven near the first core.
The region where the outflow appears in Figure~\ref{fig:5}{\it h} corresponds to the low-plasma beta region ($\betap \lesssim 0.1$) shown in Figure~\ref{fig:5}{\it i}.

\subsubsection{Outflow and Jet formation}
\begin{figure}
\begin{center}
\includegraphics[width=160mm]{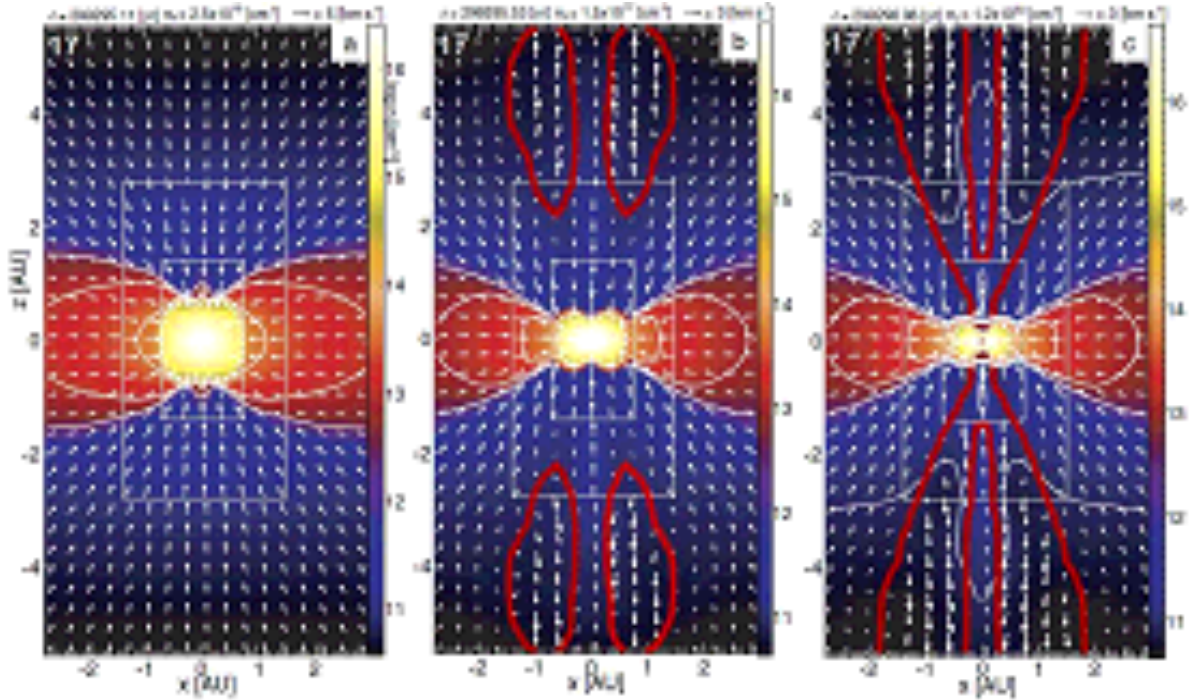}
\caption{
Time sequence of the outflow driven from the first core for model MR.
The density (color-scale) and velocity distribution (arrows) on the cross-section in the $y=0$ plane are plotted in each panel.
Panels (a) through (c) are snapshots at the stages of 
(a) $n_c =2.5 \times 10^{15} \cm$ ($l=17-19$), \ 
(b) $ 1.5 \times 10^{17} \cm$ ($l=17-19$), and \ 
(c) $ 1.2 \times 10^{23} \cm$ ($l=17-19$). 
White-dotted lines indicate the first core.
Red-solid lines indicate the border between the outflow and accretion flow (i.e., contour of $v_z$ = 0).
The elapsed time from the initial ($t$), density at the center of the cloud ($\nc$), arrow scale are denoted in each panel.
}
\label{fig:6}
\end{center}
\end{figure}

Figure~\ref{fig:6} shows the evolution of outflow for model MR.
Figures~\ref{fig:6}{\it a}--{\it c} show the density and velocity distribution around the first core on the $y=0$ plane.
The gas accretes onto the first core (white dotted lines) in the whole region outside the first core (Fig.~\ref{fig:6}{\it a}), while the gas in the regions above and below the first core (inside the red lines) outflows from the central region  (Fig.~\ref{fig:6}{\it b}).
In the ideal MHD regime, the outflow in the collapsing cloud appears in close proximity to the first core \citep{tomisaka02,matsu04,machida04,machida05b,banerjee06}.
However, as shown in Figure~\ref{fig:6}{\it b}, the outflow appears far from the first core in the non-ideal MHD regime because the magnetic field near the first core is not well coupled with the neutral gas due to Ohmic dissipation.
Thus, the torsional alfv\'en wave caused by the rotation of the first core is not effectively transferred. 
However, the torsional alfv\'en wave generated far from the first core can maintain itself, because the gas density is low and  Ohmic dissipation is less effective in these regions.
As the cloud collapses, the outflow can be driven even near the first core, as shown in Figure~\ref{fig:6}{\it c}, because in these regions, the gas density gradually decreases and  Ohmic dissipation becomes less effective over time, as shown in Figures~\ref{fig:5}{\it b}, {\it e}, and {\it h}. 
Finally, the outflow is anchored by the first core, as seen in the ideal MHD calculations.
At the end of the calculation, the outflow has a maximum speed of $3.2\km$, which is comparable to the results of ideal MHD calculations \citep{tomisaka02}.
Therefore, the outflow in our calculation (the non-ideal MHD calculation) differs from that of the ideal MHD calculation in the early phase, while the outflow features are almost the same (e.g., the shape and speed of the outflow) for both the non-ideal and ideal MHD calculations in the later phase.

\begin{figure}
\begin{center}
\includegraphics[width=160mm]{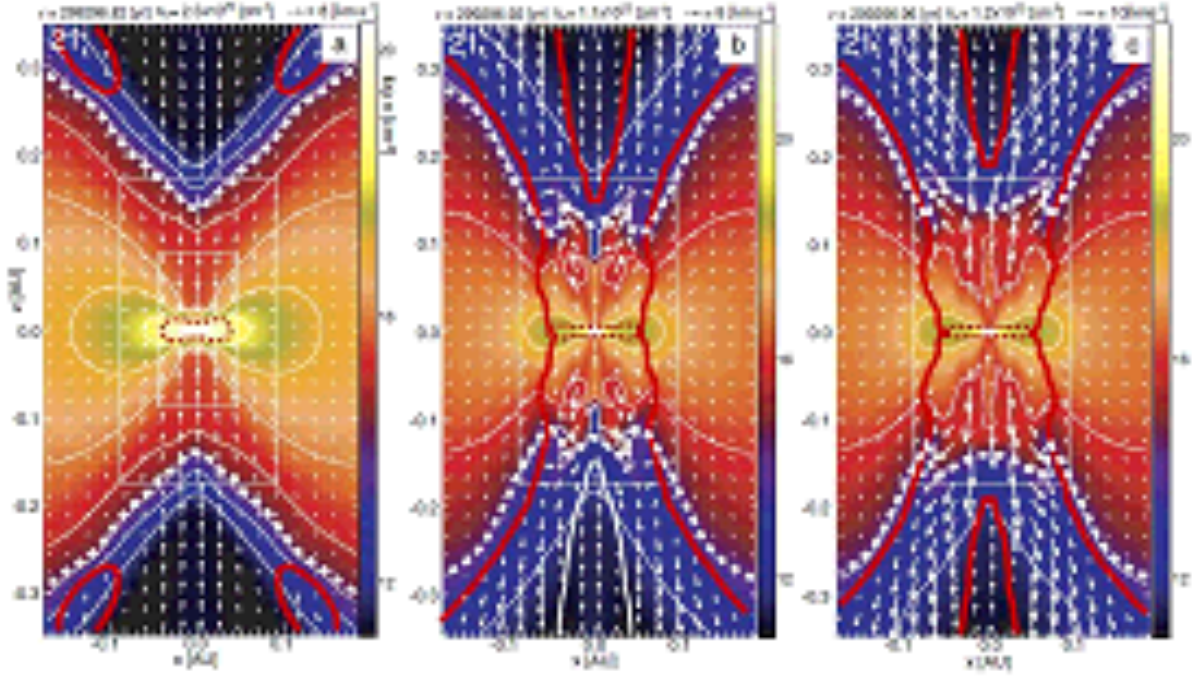}
\caption{
Time sequence of the jet outflow driven from the protostar for model MR.
The density (color-scale) and velocity distribution (arrows) on the cross-section in the $y=0$ plane are plotted in each panel.
Panels (a) through (c) are snapshots at the stages of 
(a) $n_c =2.5 \times 10^{20} \cm$ ($l=17-19$), \ 
(b) $ 1.1 \times 10^{22} \cm$ ($l=17-19$), and \ 
(c) $ 1.2 \times 10^{23} \cm$ ($l=17-19$).
The first core and second core (or protostar) are shown by white- and red-dotted lines, respectively.
Red-thick lines indicate the border between the jet and accretion flow (i.e., contour of $v_z$ = 0).
The elapsed time from the initial ($t$), density at the center of the cloud ($\nc$), arrow scale are denoted in each panel.
}
\label{fig:7}
\end{center}
\end{figure}

Figure~\ref{fig:7} shows the evolution of a jet driven from the protostar.
In this figure, the first core and protostar (i.e., the second core) are shown as dotted white and red lines, respectively.
The protostar has a disk-like structure, while the first core has a concave shape (Fig.~\ref{fig:7}{\it a}).
The gas flows out from the regions above and below the protostar (Fig.~\ref{fig:7}{\it b}).
Figure~\ref{fig:7}{\it b} shows horn-like structures at $z=\pm 0.1$\,AU, which are caused by  strong mass ejection.
This panel shows that only the gas located in the regions just above and below the protostars flows out from the protostar, while a large fraction of the gas accretes onto the protostar.
At the end of the calculation, a strong jet with a maximum velocity of $17.2\km$ is driven from the protostar, as shown in Figure~\ref{fig:7}{\it c}.

\begin{figure}
\begin{center}
\includegraphics[width=160mm]{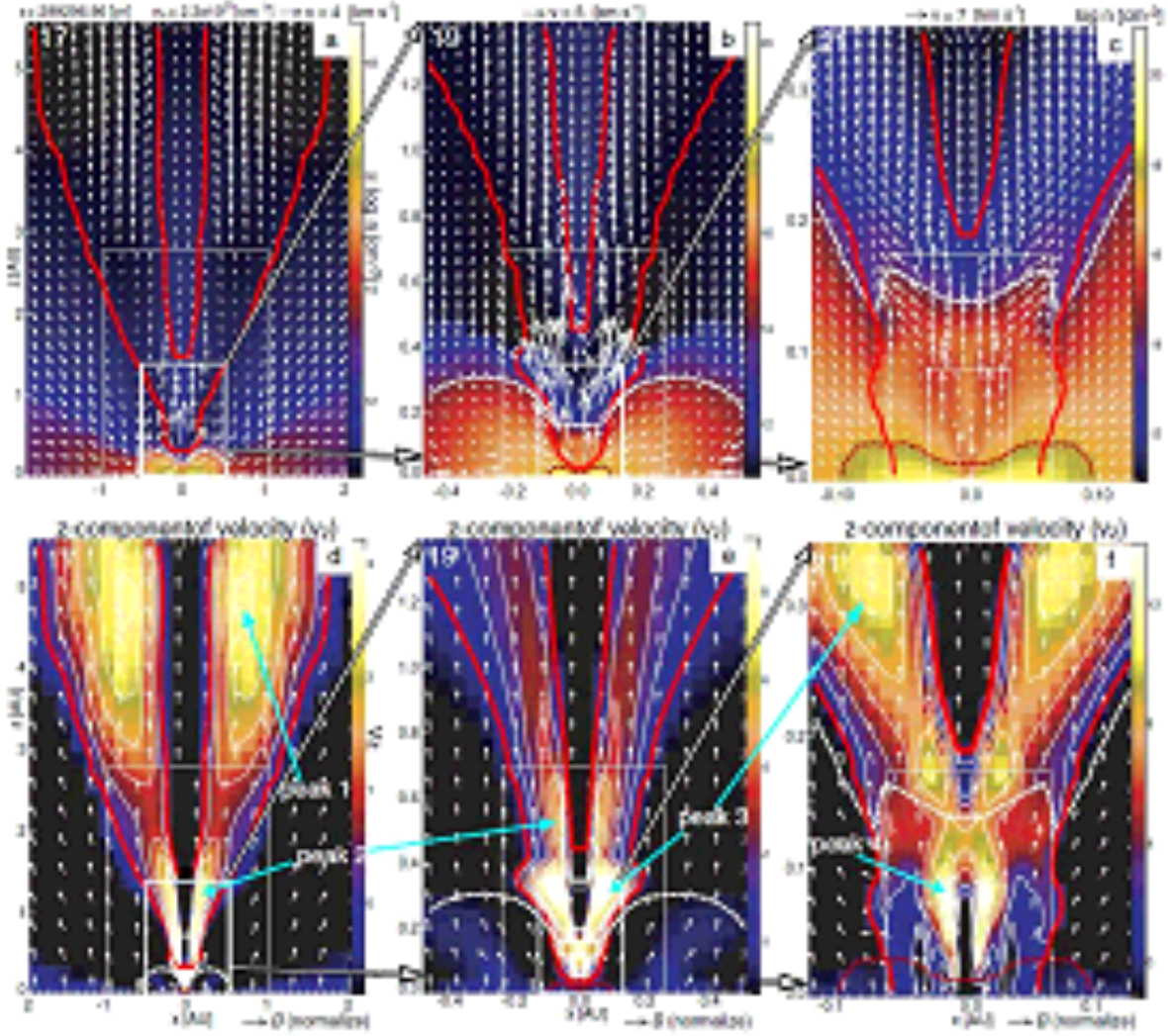}
\caption{
Upper panels: the density (color-scale and white contours) and velocity distribution (arrows) on the cross-section in the $y=0$ plane for model MR at the same epoch as Fig.~\ref{fig:6} {\it c}.
Lower panels: the vertical component of the velocity ($v_z$; color-scale and contours) and magnetic field (arrows) with the same scale as each upper panel.
The characters of `peak' indicate the position where $v_z$ has a local peak.
The first core and second core (or protostar) are shown by the white- and red-dotted lines, respectively.
Red-thick lines indicate the border between the outflow and accretion flow (i.e., contour of $v_z$ = 0).
}
\label{fig:8}
\end{center}
\end{figure}

Figures~\ref{fig:8}{\it a}--{\it c} (upper panels) show the density (color and contours) and velocity distributions (arrows) around the protostar for model MR.
The velocity of $z$-component ($v_z$; color and contours) and magnetic field (arrows) corresponding to each upper panel of Figure~\ref{fig:8} are plotted in Figures~\ref{fig:8}{\it e}--{\it f} (lower panels).
The first core and protostar (i.e., second core) are plotted as dotted white and red lines, respectively.
The outflow extends from the region just above the first core to the outside of the  $l=17$ grid boundary (Fig.~\ref{fig:8}{\it a}).
In model MR, the outflow extends to 7\,AU at the end of the calculation.

There are two velocity peaks in Figure~\ref{fig:8}{\it d}.
The upper one (peak 1) has a velocity of $v_z \simeq 2.5\km$ at $(x, z) \simeq (\pm 0.7$\,AU, 5\,AU), while the lower (peak 2) has a velocity of $v_z \simeq 3\km$ at $(x, z) \simeq (\pm 0.2$\,AU, 1\,AU).
Figures~\ref{fig:8}{\it b} and {\it e} provide close-up views of Figures~\ref{fig:8}{\it a} and {\it d}.
In Figure~\ref{fig:8}{\it e}, strong flows (peak 3) appear near the surface of the first core [($x, z) \simeq $ ($\pm0.1$\, AU, 0.3$\,$AU)]. 
Figures~\ref{fig:8}{\it c} and {\it f} show close-up views of Figures~\ref{fig:8}{\it b} and {\it e}.
A strong jet is driven from the region above the protostar in Figure~\ref{fig:8}{\it c},  corresponding to peak 4 in Figure~\ref{fig:8}{\it f}.
This peak (peak 4) has a velocity of $v_z \simeq 15\km$ at $(x, z) \simeq (\pm 0.02$\,AU, 0.08\,AU).
The lower panels of Figure~\ref{fig:8} show four velocity peaks (peak 1--peak 4).
These peaks imply that speeds (strength) of the outflow and jet change with time, because they represent a history of the outflow and jet.
We confirmed that peaks 1 and 2 appear before protostar formation, and peaks 3 and 4 appear after protostar formation.
We also confirmed that each peak is related to the oscillation of each core.
For example, peaks 1 and  2 appear every time the first core oscillates.
It is, therefore, considered that the outer peaks  (peaks 1 and 2) originate from the first core, while the inner peaks  (peaks 3 and 4) originate from the protostar.
Thus, in our definition, the outer peaks correspond to the outflow, while the inner peaks correspond to the jet.

\begin{figure}
\begin{center}
\includegraphics[width=160mm]{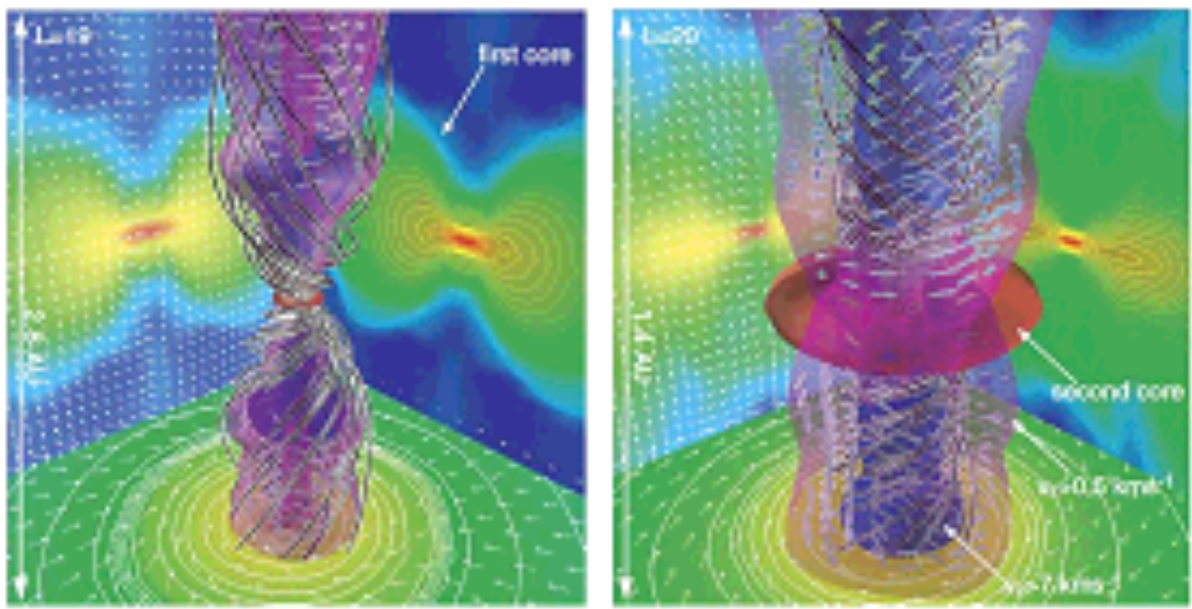}
\caption{
Bird's-eye view of model MR at the same epoch as Fig.~\ref{fig:6}{\it c} and Fig.~\ref{fig:7}{\it c}, but with different scales of $L=2.8$\,AU (left), and $L=1.4$\,AU (right).
The structure of high-density region ($\rho > 0.1\rho_{\rm c}$; red iso-density surface),  and magnetic field lines (black-and-white streamlines) are plotted in each panel.
The structures of the jet ($v \gtrsim 7\km$) and outflow ($v \gtrsim 0.5\km$) are shown by purple and blue iso-velocity surfaces, respectively.
The density contours (false color and contour lines), velocity vectors (thin arrows) on the mid-plane of $x=0$, $y=0$, and $z=$0 are, respectively, projected in each wall surface.
}
\label{fig:9}
\end{center}
\end{figure}

Figure~\ref{fig:9} shows the structure of the outflow and jet, and the configuration of the magnetic field lines.
It also shows the shapes of the first core (left panel; the projected density contours on the wall) and the protostar (right panel; the red isosurface).
The purple and blue surfaces in Figure~\ref{fig:9} indicate the iso-velocity surface of $v_z = 5 \km$ and $v_z = 0.5 \km$, respectively.
The flow inside the purple iso-velocity surface has a velocity of $v > 0.5\km$ (low-velocity component; LVC),  while the flow inside the blue iso-velocity surface has a velocity of $v>5\km$ (high-velocity component; HVC).
The HVC is enclosed by the LVC.
The LVC flow is mainly driven from the first core, while the HVC flow is mainly driven from the protostar.
The LVC and HVC are strongly coiled by the magnetic field lines anchored to the first core and protostar, respectively.

In model MR, the protostar has a mass of $4.52\times10^{-3}\msun$ and a radius of $1.07\rsun$, while at the end of the calculation, the mass flowing out from the protostar is $M_{\rm out}=1.42\times 10^{-3}\msun$.
At the same epoch, the protostar and outflow gas (outflow and jet) have angular momenta of $J_{\rm core} = 1.53\times 10^{48}\gjcm$ and $J_{\rm out} = 1.93 \times 10^{48}\gjcm$, respectively.
Therefore, the specific angular momenta of the protostar  and outflow gas (i.e., the outflow and jet) are $j_{\rm core} = 1.67\times 10^{17}\jcm$ and $j_{\rm out} = 6.82 \times 10^{17}\cm$, respectively.

\subsection{Cloud Evolution with Rapid Rotation}
\subsubsection{First Core Formation}
Figure~\ref{fig:10} shows the cloud evolution for model FR after the first core formation.
Model FR has a parameter of $\omega= 0.3$.
The ratio of the rotational to gravitational energy for model FR is $\beta_0 = 0.3$, which is $10^4$ times larger than that for the initial stage of model SR(Table~\ref{table:init}).
The first core forms at $\nc \simeq 8 \times 10^{12}\cm$, and has a mass of $M\simeq 0.026 \msun$ and a radius of $R\simeq 5.68$\,AU at its formation epoch.
The mass and size of the first core for model FR are larger than those for models SR and MR because the cloud is rotating rapidly.
The cloud in model FR has $\Omega_{\rm c}/(4\pi G \rhoc)^{1/2} \simeq 0.2$ in the initial stage, and maintains this value for $\nc \lesssim 10^{12}\cm$, as shown in Figure~\ref{fig:2}{\it d}, which indicates that the centrifugal force significantly affects the formation of the first core.
Owing to rapid rotation, the cloud has a disk-like structure near the center, as shown in Figure~\ref{fig:10}{\it a}.
The oblateness of model FR increases rapidly, as shown by the dotted line in Figure~\ref{fig:2}{\it a}, and saturates $\ob \simeq 5-6$ for $\nc \simeq 10^7\cm$.
In this model, the first core has a disk-like shape at its formation epoch.
Then, the disk-like first core deforms to a ring, and the ring fragments into several pieces at $\nc \simeq 3\times 10^{14}\cm$.
\citet{machida04,machida05b} showed that fragmentation occurs in the rapidly rotating cloud. 
Since we focus on the mechanism of the outflow and jet, we stopped the calculation when fragmentation occurred.
We will discuss the fragmentation process of the first and second cores in a subsequent paper.

\subsubsection{Outflow formation}
\begin{figure}
\begin{center}
\includegraphics[width=160mm]{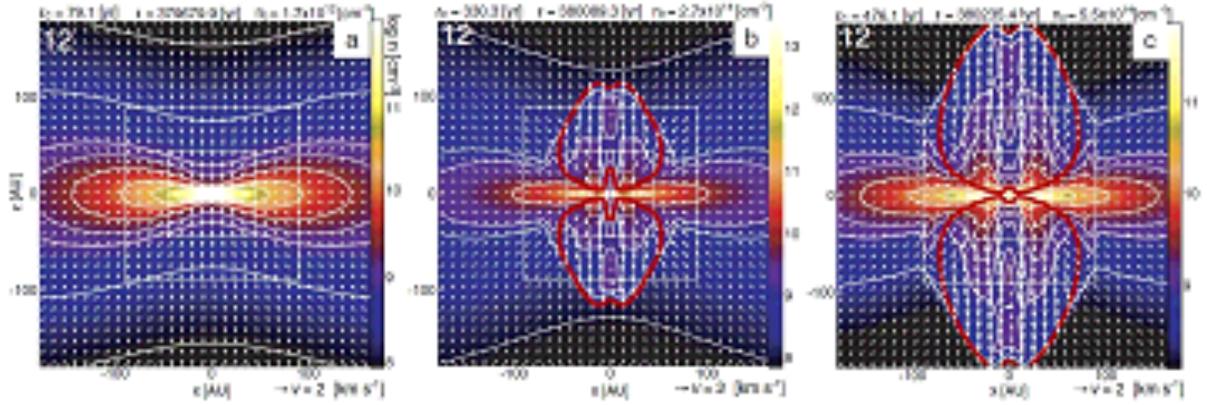}
\caption{
Time sequence of  model FR. 
The density (color-scale) and velocity distribution (arrows) on the cross-section in the $y=0$ plane are plotted.
Panels (a) through (c) are snapshots at the stages of 
(a) $n_c =1.7 \times 10^{12} \cm$ ($l=12-14$), \ 
(b) $ 2.7 \times 10^{14} \cm$ ($l=12-16$), and  
(c) $ 5.5 \times 10^{14} \cm$ ($l=16-17$).
Red-thick lines indicate the border between the outflow and accretion flow (i.e., contour of $v_z$ = 0).
The elapsed time after the first core formation ($t_{\rm c}$), elapsed time from the initial ($t$), density at the center of the cloud ($\nc$), and arrow scale are denoted in each panel.
}
\label{fig:10}
\end{center}
\end{figure}

Figure~\ref{fig:10}{\it b} shows the density and velocity distribution around the center of the cloud 330.3\,yr after the first core formation.
This panel shows that outflow is driven from the center of the cloud.
The outflow extends up to 120\,AU with a maximum speed of $v_{\rm out} \simeq 2.5\km$ at this epoch.
Figure~\ref{fig:10}{\it c} shows the structure of the outflow 476.1\,yr after the first core formation.
At the end of the calculation, the outflow extends up to 200\,AU with a maximum speed of $v_{\rm out} = 3.1\km$.
Figures~\ref{fig:10}{\it b}, \ref{fig:3}, and \ref{fig:8} show that the opening angle of the outflow for model FR is larger than that of the jet for models SR and MR. 
The height-to-radius ratio of the outflow is $H/R\simeq 2.2$ (Fig.~\ref{fig:10}{\it b}) to $2.5$ (Fig.~\ref{fig:10}{\it c}) in model FR, while that of the jet is $H/R\simeq 10$ in  model SR.
In model FR, the outflow maintains its shape (i.e., $H/R \approx\,constant$) regardless of time, while the jet in model SR becomes slender along the vertical axis with time (i.e., $H/R$ increases with time).

\begin{figure}
\begin{center}
\includegraphics[width=160mm]{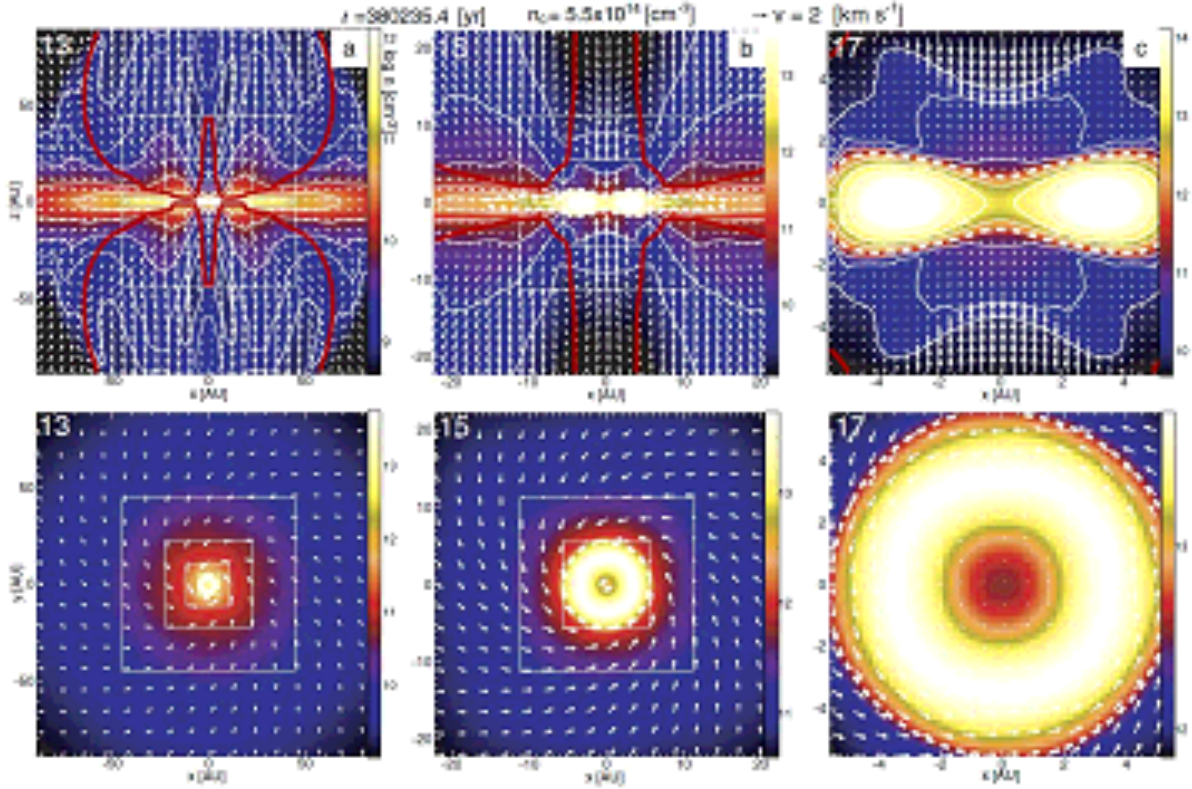}
\caption{
The density (color-scale and white contours) and velocity distribution (arrows) on the cross-section in the $y=0$ plane (upper panels) and $z=0$ plane (lower panels) at the same epoch as Fig.~\ref{fig:10}{\it c}, but with different grid levels of $l=13$ (left panels), $l=15$ (middle panels), and $l=17$ (right panels), respectively.
White-dotted lines indicate the first core.
Red-thick lines indicate the border between the outflow and accretion flow (i.e., contour of $v_z$ = 0).
}
\label{fig:11}
\end{center}
\end{figure}

Figure~\ref{fig:11} shows the density and velocity distribution on the $y=0$ (upper panel) and $z=0$ (lower panel) planes at the same epoch as Figure~\ref{fig:10}{\it c}, but with different grid scales.
Figure~\ref{fig:11}{\it c} is 16 times magnification of Figure~\ref{fig:11}{a} and shows the ripping density contours caused by the strong mass ejection from the center of the cloud. 
Since the outflow interrupts gas accretion in the regions above and below the central core, the gas accretes onto the central region only via a thin disk (Fig.~\ref{fig:11}{\it a}).
The upper panel in Figure~\ref{fig:11}{\it b} shows that the outflow is driven from the first core (dotted line).
At this epoch, the first core, which has a disk-like shape at the formation epoch, has deformed to a ring-like shape (Fig.~\ref{fig:11}{\it b} lower panel).
As shown in Figure~\ref{fig:11}{\it c}, the density of the first core exceeds $\nc \gtrsim 10^{12}\cm$, while the density of the ambient gas outside the first core is $\nc \lesssim 10^{12}\cm$.
Since  Ohmic dissipation is effective for $10^{12}\cm \lesssim \nc \lesssim 10^{15}\cm$ \citep{nakano02}, the dissipation of the magnetic field becomes effective inside the first core, while it is barely dissipated by Ohmic dissipation outside the first core.
Therefore, the magnetic field lines can be twisted just outside the first core, and the torsional Alfv\'en wave transfers to the outside, as shown in the ideal MHD calculations \citep{tomisaka02,banerjee06}.

\begin{figure}
\begin{center}
\includegraphics[width=90mm]{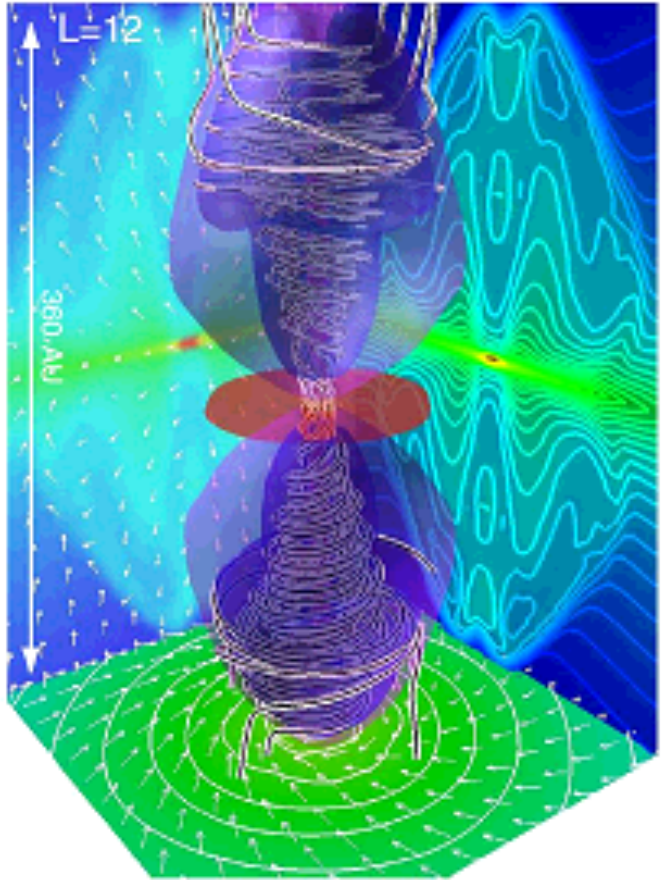}
\caption{
Bird's-eye view of model FR ($l=12$) at the same epoch as Fig.~\ref{fig:10}{\it c}.
The structure of high-density region ($n > 10^{12} \cm$; red iso-density surface), and magnetic field lines (black-and-white streamlines) are plotted.
The structure of the outflow is shown by the blue iso-velocity surface inside which the gas is outflowing from the center.  
The density contours (false color and contour lines), velocity vectors (thin arrows) on the mid-plane of $x=0$, $y=0$, and $z=$0 are, respectively, projected in each wall surface.
}
\label{fig:12}
\end{center}
\end{figure}

 Figure~\ref{fig:12} shows the shape of the outflow (blue iso-velocity surface) and the configuration of the magnetic field lines (streamlines) at the same epoch as Figure~\ref{fig:11}.
This figure shows that the magnetic field lines are strongly twisted inside the outflow, while they are loosely twisted outside the outflow. 
Due to rapid rotation, the first core forms at a lower density for model FR ($\nc \simeq 8\times 10^{12}\cm$) than for models SR and MR ($\nc \simeq 10^{14}\cm$); therefore, Ohmic dissipation is less effective outside the first core in model FR.
The magnetic field lines anchored to the rapidly rotating first core are effectively twisted, and the Lorentz force can drive the strong outflow near the first core.
We stopped the calculation for model FR at $\nc \simeq 10^{15}\cm$, because fragmentation occurred.
Thus, we cannot determine further cloud evolution for model FR.
We expect that each fragment has a small spin angular momentum because of redistribution of the angular momentum (spin and orbital angular momenta) for fragmentation \citep{machida05b}, and the magnetic field is dissipated by Ohmic dissipation in each fragment.
Thus, we expect that each fragment traces a similar evolution to models SR and MR.

In model FR, the first core has a mass of $9.42\times10^{-2}\msun$ and a radius of $5.32$\,AU, while the mass flowing out from the first core is $M_{\rm out}=6.01\times 10^{-3}\msun$ at the end of the calculation.
At the same epoch, the first core and outflow have the angular momenta of $J_{\rm core} = 8.48\times 10^{51}\gjcm$ and $J_{\rm out} = 4.95 \times 10^{50}\gjcm$, respectively.
Therefore, the specific angular momenta of the first core  and outflow are $j_{\rm core} = 4.52\times 10^{19}\jcm$, and $j_{\rm out} = 4.14 \times 10^{19}\cm$, respectively.
Thus, in model FR, the outflow largely removes the angular momentum from the first core.

\subsection{Relationship between Outflow / Jet and Initial Cloud Rotation}
In the previous sections, we showed the evolutions of three different clouds, while we investigated the cloud evolutions of 80 models in total.
As shown in \citet{machida05a,machida06a,machida07},  cloud evolution depends only on the initial angular velocity in the magnetic-force dominant clouds  $\Omega_0/B_0 < \Omega_{\rm cri}/B_{\rm cri} \equiv 0.39\, G^{1/2} \, c_s^{-1}$, where $\Omega_0$, $B_0$, and $c_{\rm s}$ are the initial angular velocity, magnetic flux density, and sound speed, respectively.
Our choice of these strongly magnetized clouds is supported by observations \citep{crutcher99,caselli02}.
From our calculations, we found the emergence conditions for the outflow and the jet fall into three categories, depending on the initial rotation rate $\omega$:
\begin{enumerate}
\renewcommand{\labelenumi}{(\arabic{enumi})}
\item  $\omega \gtrsim 0.01$, both outflow and jet can be driven from each core, 
\item  $0.002 \lesssim \omega \lesssim 0.01$, only a jet can be driven from the protostar, and 
\item  $\omega\lesssim 0.002$, neither outflow nor jet can be driven from each core. 
\end{enumerate}  
The outflow always appears when condition (1) is satisfied.
However, the jet does not appear in some models even when condition (1) or (2) is satisfied.
Thus, condition (1) is the necessary and sufficient condition for driving  outflow, while condition (1) and (2) is the necessary condition for driving  jet.
For example, although 24 models satisfy condition (2), the jet appeared only in nine models.
Although we followed the evolution of the accreting protostar up to  $\sim$200 CPU hours for each model, we did not observe any sign of the jet in those nine models.
After the protostar is formed, the magnetic field becomes strong in the regions above the poles of the protostar; thus, the plasma beta becomes $\beta \ll 1$ in these regions.
Therefore, it is difficult to calculate for a long period after protostar formation because the time step becomes significantly short owing to the increased Alfv\'en speed.
Therefore, we could not determine the reason the jet does not appear in some models.
We could not determine if the jet would never appear in those models or if it might appear much later.
To clarify the long-term evolutions, we need to develop a calculation method with implicit time integration.

\section{Discussion}
\subsection{Driving Mechanism of Outflow and Jet}
In this paper, for convenience, we called the flows driven from the first core the outflow, and the flows  driven from the protostar (i.e., the second core) the jet.
In model SR, no outflow appears because the initial cloud rotates slowly, and the rotation rate is not sufficiently amplified at the first core formation epoch.
On the other hand, we cannot observe the jet driven from the protostar in model FR because we stopped calculation once it showed fragmentation.
In model MR, both the outflow and the jet appear.
In this section, to investigate the driving mechanism of the outflow and the jet, we focus on the flows that appeared in models SR (the jet) and FR (the outflow).
We do not discuss the flows in model MR  because it is difficult to separate the jet from the outflow as shown in Figures~\ref{fig:6}--\ref{fig:9}.

At the end of the calculation, the jet has a well-collimated structure with high speed ($v_{\rm jet}\simeq 30\km$) and a large height-to-radius ratio of $H/r\gtrsim 20$ .
The $H/r$ ratio of the jet increases with time, as shown in \S 3.1.2 and 3.2.2.
The outflow, however, has a wide opening angle with slow speed ($v_{\rm out}\simeq 3\km$), and expands outwardly, maintaining the height-to-radius ratio of $H/r\simeq 2-2.5$.
The difference of the speeds between the jet and outflow can be understood from the difference in the gravitational potential.
In model SR, the protostar has a mass of $M\simeq 4.04\times 10^{-3}\msun$ and is a radius of $R\simeq 1.07\rsun$ at the end of the calculation.
The Kepler speed corresponding to the mass and radius is  $v_{\rm Kepler} = 26\km$, which is comparable to the speed of the jet in model SR ($v_{\rm jet}\simeq 30\km$).
On the other hand, at the end of the calculation for model FR, the first core has a mass of $M\simeq 9.42 \times 10^{-2}\msun$ and a radius of $R\simeq 5.32$\,AU.
Thus, the Kepler speed of the first core is $v_{\rm Kepler} = 3.96\km$, which is comparable to the speed of the outflow ($v_{\rm out}\simeq 3\km$).
As a result, both the jet and the outflow have flow speeds similar to the Kepler speed of their respective cores.

\subsubsection{Jet Driven from the Protostar}
\begin{figure}
\begin{center}
\includegraphics[width=160mm]{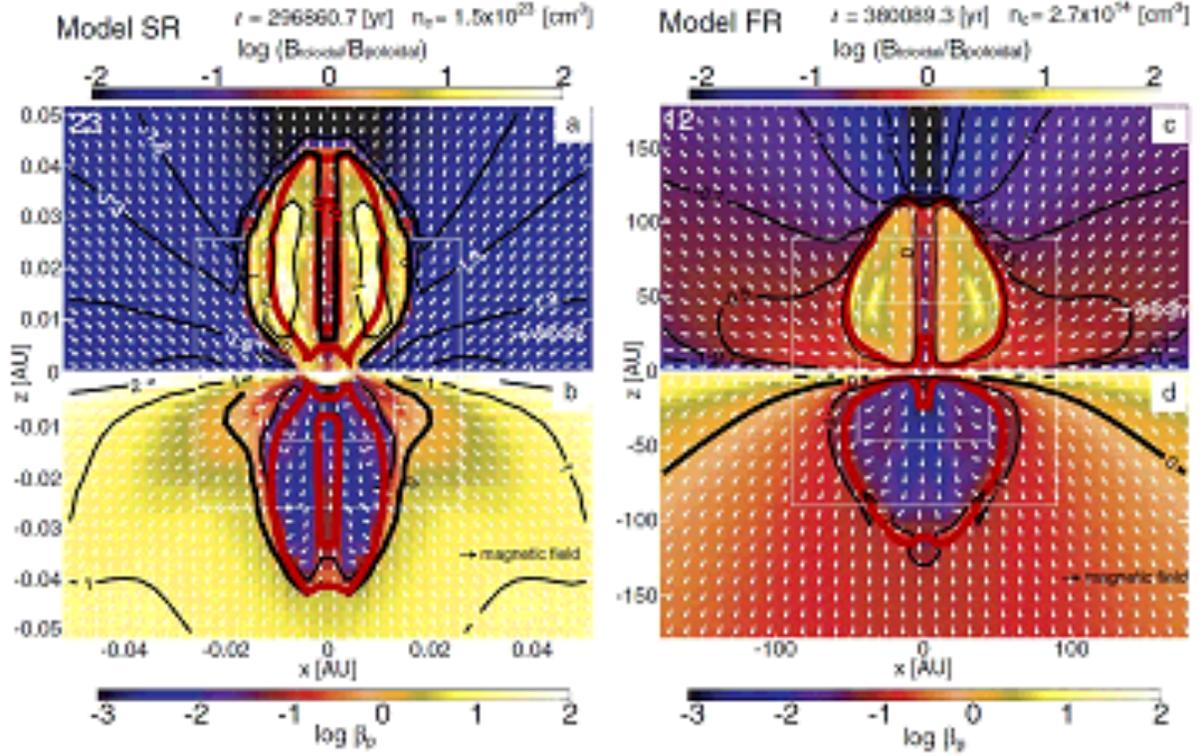}
\caption{
The ratio of the toroidal to  poloidal component of the magnetic field ($B_{\rm toroidal}/B_{\rm poloidal}$, color and contours) and velocity vector (arrows) are plotted in $z>0$ region for models with initially slow rotation model SR (left panels), and rapidly rotation FR (right panel).
The plasma beta ($\betap$; color and contours) and magnetic field (arrows) are plotted in $z<0$ region for the same models.
Red thick lines indicate the border between the outflow and accretion flow [i.e., the shapes of the jet (left panel) and outflow (right panel)].
}
\label{fig:13}
\end{center}
\end{figure}

The difference in the degree of collimation  between the outflow and the jet (the well-collimated structure of the jet  and the wide opening angle of the outflow) can be understood from the difference in their driving mechanisms and the configuration of the magnetic field lines around their drivers (i.e., the fist core and protostar).
Figure~\ref{fig:13} shows the shapes of the jet (model SR; left panels) and outflow (model FR; outflows) with thick red lines indicating the border between the accretion and the jet  or outflow. 
This shows that the collimation of the jet (left panel) is stronger than that of the outflow (right panel).

First, we discuss the collimation and driving mechanism of the jet.
The upper panel of Figure~\ref{fig:13} show the ratio of the toroidal to poloidal components of the magnetic field ($B_{\rm toroidal}/B_{\rm poloidal}$) at each mesh point around the outflow (model SR; left) and the jet (model MR; right). 
In Figure~\ref{fig:13}{\it a}, the toroidal field dominates the poloidal field inside the jet ($B_{\rm toroidal}/B_{\rm poloidal} \simeq 10$), while the toroidal field barely exists outside the jet ($B_{\rm toroidal}/B_{\rm poloidal} \lesssim 0.01$).
In model SR, the magnetic field is dissipated by Ohmic dissipation, and decoupled from the neutral gas inside the first core (or outside the second core), as shown in Figure~\ref{fig:1}{\it i}.
Thus, the magnetic field lines are relaxed  by the magnetic tension force and are almost straight  ($B_z \gg B_r, B_\phi$ in the cylindrical coordinates), because magnetic field lines move freely, irrespective of the neutral gas.
Even after the magnetic field is well-coupled with the neutral gas for $\nc \gtrsim 10^{15}\cm$, the straight magnetic field lines are distributed around the protostar, which reflects past Ohmic dissipation.
Therefore, the vertical component of the magnetic field ($B_z$) is dominant outside the protostar.
The lower panels in Figure~\ref{fig:13} show the plasma beta around the jet (left; model MR) and the outflow (right; model FR).
Figure~\ref{fig:13}{\it b} shows that the magnetic field is weak ($\betap \simeq 10$) outside the jet (outside the red line), while it is very strong ($\betap\lesssim 0.01$) inside the jet (inside the red line), compared to the thermal pressure.
The sharing motion between the protostar and ambient gas amplifies the magnetic field inside the jet,  since the formed protostar rapidly rotates because the magnetic braking is less effective, explained in \S3.1.
In general, when the magnetic field is weak around the driver, it is considered that the magnetic pressure gradient force dominates the magnetocentrifugal force in terms of driving the flow \citep{uchida85,tomisaka02}.

\begin{figure}
\begin{center}
\includegraphics[width=140mm]{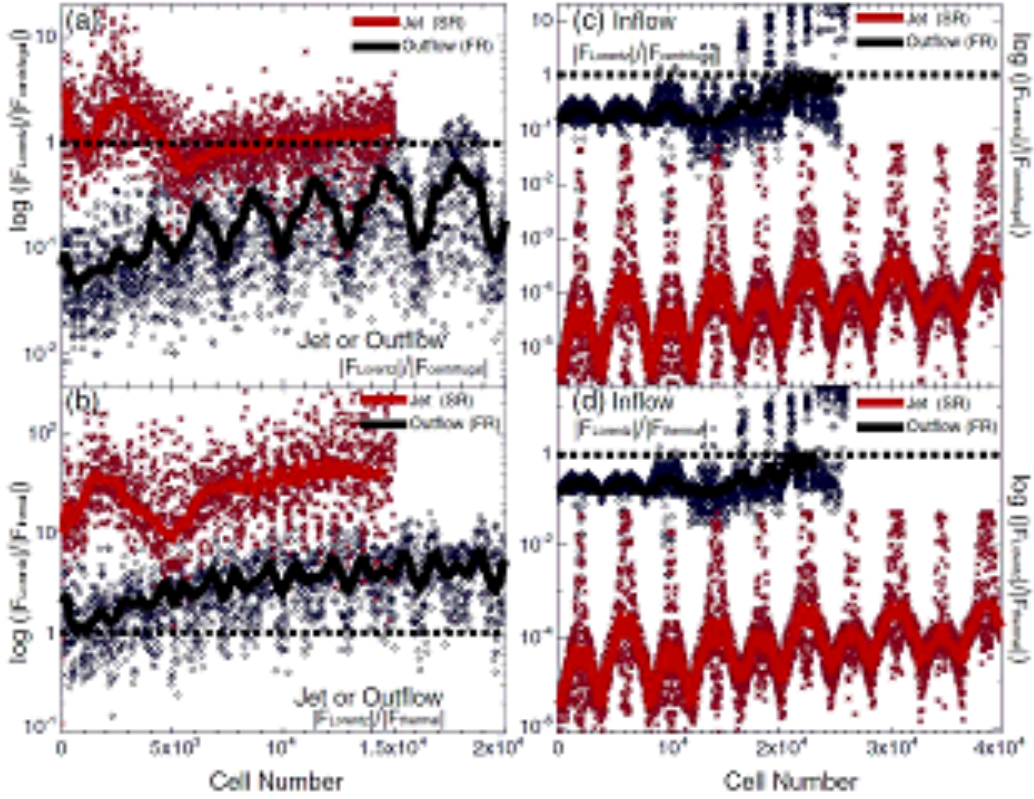}
\caption{
The ratio of each force for models SR and FR at the end of the calculation.
Panel {\it a}: Ratio of the Lorentz  to the centrifugal forces ($F_{\rm Lorentz}/F_{\rm centrifugal}$) at each mesh point in the region with $v_{z} > 0.1\km$ (i.e., the outflow and jet region).
Panel {\it b}: The ratio of the Lorentz to the thermal pressure gradient forces ($F_{\rm Lorentz}/F_{\rm thermal}$) in the region with $v_{z} > 0.1\km$ (i.e., the outflow and jet region).
Panel {\it c}: The ratio of the Lorentz to the centrifugal forces ($F_{\rm Lorentz}/F_{\rm centrifugal}$) in the region with $v_{z} < 0.1\km$ (i.e., the inflow region).
 and 
Panel {\it d}: The ratio of the Lorentz to the centrifugal forces ($F_{\rm Lorentz}/F_{\rm centrifugal}$) in the region with $v_{z} < 0.1\km$ (i.e., the inflow region).
The horizontal axis means the number of the cell.
The dots are the data from models FR (black dots), and SR (red dots), respectively.
The thick lines are average of adjacent 100 cells for models FR (black lines), and SR (red lines), respectively.
The horizontal-dotted lines indicate the value at which each force equals.
}
\label{fig:14}
\end{center}
\end{figure}

To investigate the driving mechanism of the jet and the outflow, we calculate the Lorentz ($F_{\rm Lorentz}$), centrifugal ($F_{\rm centrifugal}$), and thermal pressure gradient ($F_{\rm pressure}$) forces parallel to the poloidal component of the magnetic field ($B_{\rm p}$).
When the self-gravity term in equation~(\ref{eq:eos}) is ignored, the equation of motion can be rewritten as
\begin{equation}
\rho \dfrac{\partial \vect{v}}{\partial t} = -\rho (\vect{v} \cdot \vect{\nabla}) -\nabla P - \dfrac{1}{4\pi}\left[ \vect{B}\times (\vect{v}\times \vect{B}) \right].
\label{eq:eom2}
\end{equation}
We consider the dot product between each term in the left-hand side in the equation~(\ref{eq:eom2}) and the poloidal component of the magnetic field ($\vect{B_{\rm p}}$).
Thus, each force parallel to the $\vect{B_{\rm p}}$ is written as
\begin{eqnarray}
F_{\rm centrifugal}  &\equiv&   \vert \rho (\vect{v} \cdot \vect{\nabla}) \cdot \vect{e_{\rm p}} \vert, \\
F_{\rm thermal}      &\equiv&   \vert \nabla P  \cdot \vect{e_{\rm p}} \vert, \\
F_{\rm Lorentz}      &\equiv&    \dfrac{1}{4\pi} \vert \vect{B}\times (\vect{v}\times \vect{B}) \cdot \vect{e_{\rm p}} \vert, 
\end{eqnarray}
where $\vect{e_{\rm p} \equiv \vect{B_{\rm p}/\vert\vect{B_{\rm p}} \vert }}$.
In the left panels of Figure~\ref{fig:14}, we plot the ratio of the Lorentz to the centrifugal force ($F_{\rm Lorentz}/F_{\rm centrifugal}$; Fig.~\ref{fig:14}{\it a}), and the ratio of the Lorentz to the thermal pressure gradient force ($F_{\rm Lorentz}/F_{\rm thermal}$; Fig.~\ref{fig:14}{\it b}) in the outflow region at each mesh point, where the outflow region is defined as meshes with $v_{\rm z} > 0.1\km$ for $z>0$. 
The right panels (Fig.~\ref{fig:14}{\it c} and {\it d}) show the plots for the same ratios of $F_{\rm Lorentz}/F_{\rm centrifugal}$, and $F_{\rm Lorentz}/F_{\rm thermal}$ in the accreting region ($v_{\rm z}<0.1\km$ for $z>0$).
In Figure~\ref{fig:14}, red dots and lines are data from the region around the jet in model SR, while  black dots and lines are data from the region around the outflow in model FR.
In Figure~\ref{fig:14}, left panels show that, inside the jet (red dots and line), the Lorentz force is equivalent to the centrifugal force ($F_{\rm Lorentz}/F_{\rm centrifugal} \simeq 1$; Fig.~\ref{fig:14}{\it a}), while the Lorentz force is about $\sim30-40$ times stronger than the thermal pressure gradient force ($F_{\rm Lorentz}/F_{\rm centrifugal} \simeq 30-40$; Fig.~\ref{fig:14}{\it b}).
In the accreting region around the jet (red dots and lines  in Fig.~\ref{fig:14}{\rm c} and {\rm d}), the Lorentz force is much weaker than both  the centrifugal ($F_{\rm Lorentz}/F_{\rm centrifugal}\simeq 10^{-5}-10^{-4}$) and thermal pressure gradient forces  ($F_{\rm Lorentz}/F_{\rm pressure}\simeq 10^{-6}-10^{-4}$).
Thus, the Lorentz force is not effective in the accreting regions.

In model SR, a rapidly rotating protostar forms because the magnetic field around the center of the cloud is weak and magnetic braking is less effective.
Thus, the magnetic field lines anchored to the protostar make a strong toroidal field for the rapid rotation of the protostar.
The toroidal field caused by the protostar transmits in the vertical direction as the torsional alfv\'en wave, generating a large magnetic pressure gradient along the vertical axis.
The jet is considered to be driven by this strong magnetic pressure gradient force.
In general, the jet has  good collimation when the magnetic pressure gradient force is dominant ($B_{\rm toroidal} \gg B_{\rm poloidal}$) for driving the flow, because the magnetic field lines are pinched by the toroidal field, as shown by \citet{tomisaka02}.
Furthermore, the jet in model SR is guided by the straight configuration of the magnetic field lines ($B_z \gg B_r, B_{\rho}$) outside the jet region.
Thus, the well-collimated jet is caused by both the driving mechanism of the magnetic pressure gradient and the straight configuration of the magnetic field lines around the protostar.

\subsubsection{Outflow Driven from the First Core}
Figure~\ref{fig:13} shows the ratio of the toroidal to poloidal components of the magnetic field ($B_{\rm toroidal}/B_{\rm poloidal}$; Fig.~\ref{fig:13}{\it c}) and the plasma beta ($\betap$; Fig.~\ref{fig:13}{\it d}) around the outflow for model FR.
The toroidal field is $\sim 3-5$ times stronger than the poloidal field ($B_{\rm toroidal}/B_{\rm poloidal} \simeq 3-5$) inside the outflow (inside the red line).
However, this ratio  is smaller than that of the jet ($B_{\rm toroidal}/B_{\rm poloidal} \simeq 10$; model SR) because the rotation rate in model FR, which causes the toroidal field, is small around the first core, due to effective magnetic braking. 
Outside the outflow, however, the poloidal field is slightly larger than the toroidal field ($B_{\rm toroidal}/B_{\rm poloidal} \simeq 0.3$).
Figure~\ref{fig:13}{\it c} shows that the magnetic energy is larger than the thermal energy inside the outflow ($\betap\lesssim 0.01$), while the magnetic energy is comparable to or slightly weaker than the thermal energy outside the outflow ($\betap\sim0.1-1$).
The gas around the first core has a density range of $10^8\cm \lesssim \nc \lesssim 10^{12}\cm$ in model SR (Fig.~\ref{fig:10}).
Since Ohmic dissipation is ineffective in this region, the magnetic energy becomes comparable to the thermal energy as demonstrated by \citet{machida05a,machida06a}.
After the first core formation, the toroidal field can be amplified because the rotational timescale is shorter than the collapsing timescale inside the first core, and the magnetic field lines begin to twist \citep{machida06d}.
Therefore, the magnetic energy inside the outflow (or near the first core) becomes larger than that outside the outflow (or  far from the first core).

As shown in Figures~\ref{fig:14}{\it a} and {\it b}, the centrifugal force is dominant over the Lorentz and thermal pressure gradient forces inside the outflow ($F_{\rm Lorentz}/F_{\rm centrifugal} < 1$, and $F_{\rm Lorentz}/F_{\rm thermal} \gtrsim$ 1; black dots and lines), while it is comparable to the Lorentz force inside the jet (red dots and lines).
Outside the outflow, both the centrifugal and thermal pressure gradient forces are stronger than the Lorentz force  ($F_{\rm centrifugal}\simeq F_{\rm thermal} > F_{\rm Lorentz}$), since $F_{\rm Lorentz}/F_{\rm centrifugal} \simeq 0.1$ (Fig.~\ref{fig:14}{\it c}) and $F_{\rm Lorentz}/F_{\rm thermal} \simeq 0.1$ (Fig.~\ref{fig:14}{\it d}).
These results indicate that the outflow in model SR is mainly driven by the magnetocentrifugal mechanism \citep{blandford82}, which causes the loosely pinched magnetic field lines inside the outflow.
In addition, before the outflow appears, the magnetic field lines around the first core expand sideways toward the rotation axis from the center of the cloud, whose shape looks like a capital letter U or V, as shown by \citet{tomisaka02}.
Thus, the collimation of the magnetic field lines becomes worse with distance from the center, as shown in Figure~8{\it a} of \citet{machida06d}.
Since the outflow traces this configuration of the magnetic field lines, it does not have good collimation and the outflow driven from the first core has a wide opening angle because of the magnetocentrifugal driving mechanism and the configuration of the magnetic field lines. 
The former realizes the loosely pinched magnetic field lines, and the latter swells the outflow.

\subsection{Dependence on the Initial Magnetic field and Rotation Rate}
In this paper, we investigated cloud evolution with same magnetic field strengths ($B_{\rm ini} = 17\mu$G) but different rotation rates [$\Omega_{\rm ini} =7 \times 10^{-16}$\,s$^{-1}$ (model SR), $7 \times 10^{-15}$\,s$^{-1}$ (model MR), and $7\times 10^{-14}$\,s$^{-1}$ (model FR)].
As shown in previous section, the speeds of the jets and outflows depend on the mass and size of their drivers.
The mass and size of the first and second cores at their formation epochs depend on the initial rotation rate.
Therefore, the speeds of the jet and outflow depend on the initial rotation rate of the cloud core.
For example, in a rapidly rotating clouds, the first core forms at lower density with larger size.
When the first core has a large size and low density, the outflow is slow because the gravitational potential is shallow.
On the other hand, a slowly rotating cloud produces a first core with smaller size and higher density.
In this case, there is high-speed outflow, owing to the deep gravitational potential of the first core.
Thus, the outflow has higher speed  in a more slowly (rapidly) rotating cloud.
In addition, a cloud with an extremely low rotation rate produces no outflow because the first core has such a small angular momentum at its formation epoch that the magnetic field lines are barely twisted.
The speed of the jet, however, depends on the initial rotation rate only slightly because the angular momentum around the center of the cloud converges to a certain value by the time the protostar (i.e., second core) is formed \citep{machida07}.
Thus, the variation caused by the rotation of initial cloud is not a factor affecting the time of the protostar formation epoch.
Note that the properties of the jet and outflow might possibly change much later because we followed the evolution of these flows for only a short period.

Does the difference of magnetic field strength in the initial cloud affect the properties of the jet and outflow? 
\citet{machida05a,machida06a} showed that the evolution of the molecular cloud depends only on the ratio of the angular velocity to the magnetic flux density of the initial cloud ($\Omega_{\rm ini}/B_{\rm ini}$).
Initial differences in the distribution of the density, velocity, and magnetic field hardly affect cloud evolution, especially once the central region becomes adiabatic  ($\nc \gtrsim 10^{11}\cm$).
Cloud evolution is controlled by the magnetic field when $\Omega_{\rm ini}/B_{\rm ini} < 0.39\, G^{1/2} c_s^{-1}$, while it is controlled by the rotation  when $\Omega_{\rm ini}/B_{\rm ini} > 0.39\, G^{1/2} c_s^{-1}$ \citep{machida06a}.
Although we can choose arbitrary parameter of $\Omega_{\rm ini}$ and $B_{\rm ini}$, observations indicate that the magnetic energy is much larger than the rotation energy in many molecular clouds \citep{crutcher99,caselli02}, and clouds have $\Omega_{\rm ini}/B_{\rm ini} > 0.39\, G^{1/2} c_s^{-1}$ (i.e., the magnetic-force dominant clouds in \citealt{machida05a}).
In magnetic-force dominant clouds, the magnetic field converges to a specific value in the isothermal phase (the magnetic flux - spin relation; \citealt{machida05a,machida06a,machida07}).
Thus, cloud evolution does not depend much on the initial magnetic field strength.
As a result, outflows and jets with similar properties may appear in these clouds when they have the same angular velocities at their initial state.

\subsection{Duration of Driving Outflow}
In this paper, we have shown  that the outflow is driven from the first core.
After the first core formation epoch,  the collapse timescale becomes longer than the rotational timescale, the magnetic field lines are strongly twisted for the rotating first core, and outflow appears as shown in \S3.2.2.
Observations have shown  that outflows have lifetimes of $\sim6\times10^4$\,yr \citep{wu04}.
However, since we followed cloud evolution only through the very early phase, we could not determine whether outflow continues to be driven for a long time.
The spherical symmetric calculation indicates that the first core disappears $\sim1000$ yr after its formation epoch.
In model FR,  we followed cloud evolution for $\sim800$\,yr after the first core is formed.
In this model, the first core remains until the end of the calculation.
The outflow that appears just after the first core formation epoch continues to be driven from the first core.
The rotating first cores seem to have lifetimes of several thousand years or more \citep[e.g.,][]{saigo06}.
Thus, outflows driven from these cores may maintain themselves for several thousand years.
Note that it is expected that a jet will continue to be driven from the protostar once it appears because the protostar does not disappear.

After the first core disappeared, we could not determine whether the outflow continues to be driven.
\citet{saigo06} suggested that after the first core disappears, the remnant of the first core forms a disk-like structure or torus.
Outflow may continue to be driven from these objects even after the disappearance of the first core.
However, if the outflow does not continue after the first core disappears, then outflow, as shown in this paper, might be a transient phenomena, in which the outflow is driven only for $\sim1000$\,yr.
In this case, the flow from the first core, which we call outflow in this paper,  might be a special phenomenon that would be observed only in limited numbers of young stellar objects.
To understand the duration of the outflow driven from the first core, further long-term calculations, possibly with an implicit code, are necessary.
At present we conjecture that the outflow in our calculations will be driven for $\gtrsim10^4$ years.

\section{Summary}
Observation shows that the outflows have wide opening angles and low flow speeds, while jets have good collimation and high flow speeds.
Outflow has been considered to be entrained by the jet driven from a circumstellar disk around the protostar.
In this paper, we calculated the cloud evolution from the molecular cloud core to  protostar formation.
As a result of calculations, we found that two distinct flows (outflow and jet) are driven from different objects (the first, and second cores), and the features of outflow and jet (flow speed and collimation factor) were naturally reproduced.

\begin{figure}
\begin{center}
\includegraphics[width=80mm]{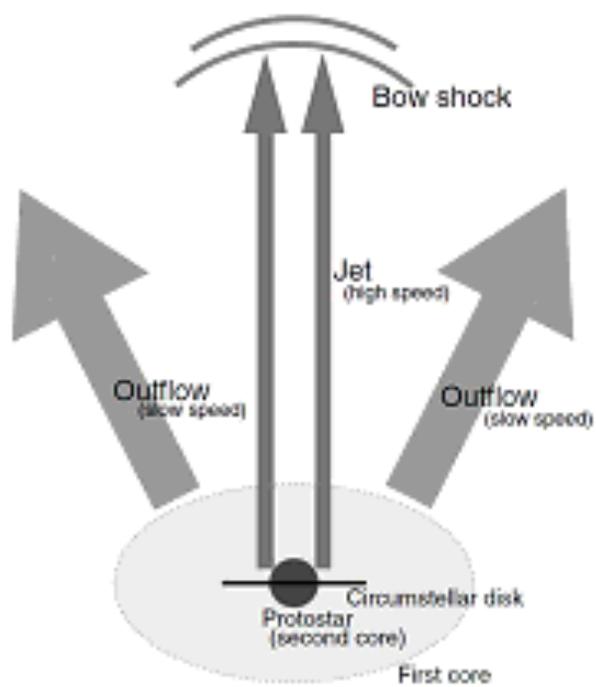}
\caption{
Schematic figure of the jet and outflow driven from the protostar and the fist core, respectively.
}
\label{fig:15}
\end{center}
\end{figure}

Our results show that the flow appearing around the first core has a wide opening angle and slow speed, while the jet appearing around the protostar  has a well-collimated structure and high speed, as shown in Figure~\ref{fig:15}.
The speed difference is caused by the difference of the depth in the gravitational potential.
The flow speed (i.e., the speed of the outflow and jet) corresponds to the Kepler speed of each object (the first and the second cores).
Because the first core has a shallow gravitational potential, its flow (outflow) is slower.
The flow (jet) driven from the protostar, which has a deeper gravitational potential, has a high speed.
In our calculations, the outflow and the jet have speeds of $v_{\rm out} \simeq 3\km$ and $v_{\rm jet}\simeq 30\km$, respectively.
These speeds are slower than those of observations. 
Typically, observed molecular outflow and optical jet have speeds of  $v_{\rm out,obs} \simeq 30\km$, and $v_{\rm jet,obs}\simeq 100\km$, respectively.
However, since the first and second cores (protostar) have mass of $M_{\rm first\,core} = 0.01\msun$ and $M_{\rm second\, core} \simeq 10^{-3}\msun$, respectively, at the end of the calculations, each core increases its mass in the subsequent gas accretion phase. 
The Kepler speed increases with the square root of the mass of the central object at a fixed radius.
When the mass of each core increases by 100 times, the Kepler speed increases 10 times.
Thus, the speed of the outflow and jet may increase by 10 times, and reach  $v_{\rm out} \simeq 30\km$ and $v_{\rm jet}=300\km$, respectively, which correspond to typical observed values.

The outflow has a wide opening angle, while the jet has a well-collimated structure.
This is caused both by the configuration of the magnetic field lines around the drivers and their driving mechanisms.
The magnetic field lines around the first core have an hourglass configuration because they converge to the cloud center  as the cloud collapses, and Ohmic dissipation is ineffective before the first core formation.
The flow appearing near the first core is mainly driven by the magnetocentrifugal wind mechanism (disk wind).
The centrifugal force is dominant in the outflow, whereas near the protostar, the magnetic field lines are a straight, and the magnetic pressure gradient mechanism is more effective for driving the jet.
The magnetic field lines are stretched by the magnetic tension force near the protostar because the magnetic field is decoupled from the neutral gas.
However, the magnetic field lines are strongly twisted in the region in close proximity to the protostar, where the magnetic field is coupled with the neutral gas again.
Thus, the strong toroidal field generated around the protostar can drive the jet, which is guided by the straight configuration of the magnetic field.

Our calculations do not completely reject the well-known concept that the observed molecular outflow is entrained by the jet, because we calculate the formation of the jet and outflow only in the early star-formation phase.
Further long-term calculations are needed to understand the mechanism of the jet and outflow in more detail.

\acknowledgments
We have greatly benefited from the discussion with ~T. Nakano and ~K. Saigo.
We also thank T. Hanawa for making a contribution to the nested grid code.
Numerical calculations were carried out with a Fujitsu VPP5000 at the Astronomical Data Analysis Center, the National Astronomical Observatory of Japan.
This work is supported by the Grant-in-Aid for the 21st Century COE "Center for Diversity and Universality in Physics" from the Ministry of Education, Culture, Sports, Science and Technology (MEXT) of Japan, and partially supported by 
the Grants-in-Aid from MEXT (15740118, 16077202, 18740113, 18740104).

\clearpage
\begin{table}   
\caption{Model parameters and calculation results}
\label{table:init}
\begin{center}
\begin{tabular}{ccccccccccccccccccc|cc}
\hline
 Model  & $\omega$ &  $B_0$ {\scriptsize [$\mu$G]} & $\Omega_0$ {\scriptsize [s$^{-1}$]}  &  $M_0${\scriptsize [$\msun$]} &$\alpha_0$ &$\beta_0$ & $\gamma_0$ & $B_{\rm f}$ {\scriptsize  (kG)} & P {\scriptsize (day)} \\
\hline
SR  & $0.003$  & 17& $7.0 \times 10^{-16}$ &7.6& 0.5& $3\times10^{-5}$ & 0.96 & 2.18 & 3.0\\
MR  & $0.03 $  & 17& $7.0 \times 10^{-15}$ &7.6& 0.5& $3\times10^{-3}$ & 0.96 & 0.40 & 2.1\\
FR  & $0.3  $  & 17& $7.0 \times 10^{-14}$ &7.6& 0.5& $3\times10^{-1}$ & 0.96 & --- & --- \\
\hline
\end{tabular}
\end{center}
\end{table}
\clearpage

\clearpage

\clearpage

\clearpage

\clearpage

\clearpage

\clearpage

\clearpage

\clearpage

\clearpage

\clearpage

\clearpage

\clearpage

\clearpage

\clearpage

\end{document}